\documentclass{vldb}

\pdfoutput=1
\usepackage{times}
\usepackage{amsmath}
\usepackage{listings}
\usepackage{color}
\usepackage{subfigure}
\usepackage{graphicx}
\usepackage{hyperref}
\usepackage{algorithm}
\usepackage[noend]{algpseudocode}
\usepackage{setspace}
\usepackage{url}
\usepackage{balance} 

\def\algospacing{0.9}

\begin{document}

\title{GraphMat: High performance graph analytics made productive}

\numberofauthors{7}

\author{
\alignauthor
Narayanan Sundaram\\
\affaddr{Intel Corporation}\\
\email{\normalsize{narayanan.sundaram@intel.com}}
\alignauthor
Nadathur Satish\\
\affaddr{Intel Corporation}\\
\email{\normalsize{nadathur.rajagopalan.satish@intel.com}}
\alignauthor
Md Mostofa Ali Patwary\\
\affaddr{Intel Corporation}\\
\email{\normalsize{mostofa.ali.patwary@intel.com}}
\and
\alignauthor
Subramanya R Dulloor\\
\affaddr{Intel Corporation}\\
\email{\normalsize{subramanya.r.dulloor@intel.com}}
\alignauthor
Satya Gautam Vadlamudi\\
\affaddr{Intel Corporation}\\
\email{\normalsize{satya.gautam.vadlamudi@intel.com}}
\alignauthor
Dipankar Das\\
\affaddr{Intel Corporation}\\
\email{\normalsize{dipankar.das@intel.com}}
}

\additionalauthors{Additional authors: Pradeep Dubey (Intel Corporation, \\
{\texttt{pradeep.dubey@intel.com}})}

\maketitle

\begin{abstract}
Given the growing importance of large-scale graph analytics, there is
a need to improve the performance of graph analysis frameworks without 
compromising on productivity.
GraphMat is our solution to bridge this gap between a user-friendly
graph analytics framework and native, hand-optimized code. GraphMat
functions by taking vertex programs and mapping them to high performance
sparse matrix operations in the backend. We get the productivity
benefits of a vertex programming framework without sacrificing performance. 
GraphMat is in C++, and we have been able
to write a diverse set of graph algorithms in this framework with the same effort compared
to other vertex programming frameworks. GraphMat performs 1.2-7X faster than high performance 
frameworks such as GraphLab, CombBLAS and Galois. It achieves better multicore scalability (13-15X on 24 cores)
than other frameworks and is 1.2X off native, hand-optimized
code on a variety of different graph algorithms. Since GraphMat 
performance depends mainly on a few scalable and well-understood sparse matrix operations, GraphMat
can naturally benefit from the trend of increasing parallelism on future hardware. 

\end{abstract}

\section{Introduction} \label{sec:intro}

Studying relationships among data expressed in the form of graphs has become increasingly important. Graph 
processing has become an important component of bioinformatics~\cite{bioinfographs}, social network analysis~\cite{Twitter10,facebook09}, traffic engineering~\cite{traffic-engg} etc. With graphs getting larger and queries getting more complex, there is a need for graph 
analysis frameworks to help users extract the information they need with minimal programming effort. 

There has been an explosion of graph programming frameworks in recent years \cite{giraph,combblas, galois, graphlab, pegasus, Seo:2013:VLDB}. All of them claim to 
provide good productivity, performance and scalability. However, a recent study has shown \cite{SIGMOD:2014} that 
the performance of most frameworks is off by an order of magnitude
when compared to native, hand-optimized code. Given that much of this
performance gap remains even when running frameworks on a single node~\cite{SIGMOD:2014}, 
it is imperative to maximize the efficiency of graph frameworks on existing hardware (in addition 
to focusing on scale out issues). GraphMat is our solution to bridge this performance-productivity gap in graph analytics.

The main idea of GraphMat is to take vertex programs and map them to a generalized sparse matrix vector multiplication operation. We get the productivity benefits of vertex programming while enjoying the 
high performance of a matrix backend. In addition, it is easy to understand and reason about, while letting users with knowledge of vertex programming a smooth transition to a high performance environment. Although other graph frameworks based on matrix operations exist (e.g. CombBLAS \cite{combblas} and PEGASUS \cite{pegasus}), GraphMat wins out in terms of both productivity and performance as GraphMat is faster and does not expose users to the underlying matrix primitives (unlike CombBLAS and PEGASUS). We have been able to write multiple graph algorithms in GraphMat with the same effort as other vertex programming frameworks. 

Our contributions are as follows: 

\begin{enumerate}
\item GraphMat is the first vertex programming model to achieve within 1.2X of native, hand-coded, optimized code on a variety of different graph algorithms. GraphMat is 5-7X faster than GraphLab \cite{graphlab} \& CombBLAS and 1.2X faster than Galois \cite{galois} on a single node.
\item %
GraphMat achieves good multicore scalability, getting a 13-15X speedup over a single threaded implementation on 24 cores. In comparison, GraphLab, CombBLAS, and Galois scale by only 2-12X over their corresponding single threaded implementations.  
\item GraphMat is productive for both framework users and developers. Users do not have to learn a new programming paradigm (most are familiar with vertex programming), whereas backend developers have fewer primitives to optimize as it is based on Sparse matrix algebra, which is a well-studied operation in High Performance Computing (HPC) \cite{SPMV}. 
\end{enumerate}

Matrices are fast becoming one of the key data structures for databases, with systems such as SciDB \cite{scidb} and other array stores becoming more popular. Our approach to graph analytics can take advantage of these developments, letting us deal with graphs as special cases of sparse matrices. Such systems offer transactional support, concurrency control, fault tolerance etc. while still maintaining a matrix abstraction. We offer a path for array processing systems to support graph analytics through popular vertex programming frontends.

Basing graph analytics engines on generalized sparse matrix vector multiplication (SPMV) has other benefits as well. We can leverage decades of research on techniques to optimize sparse linear algebra in the High Performance Computing world. Sparse linear algebra provides a bridge between Big Data graph analytics and High Performance Computing. Other efforts like GraphBLAS \cite{graphblas} are also part of this growing effort to leverage lessons learned from HPC to help big data. 

The rest of the paper is organized as follows. Section \ref{sec:motivation} provides motivation for GraphMat and compares it to other graph frameworks. Section \ref{sec:algorithms} discusses the graph algorithms used in the paper. Section \ref{sec:graphmat} describes the GraphMat methodology in detail. Section \ref{sec:results} gives details of the results of our experiments with GraphMat while Section \ref{sec:conclusion} concludes the paper.

\section{Motivation and related work} \label{sec:motivation}

Graph programming frameworks come in a variety of different programming models. Some common ones 
are vertex programming (``think like a vertex''), matrix operations (``graphs are sparse matrices''), task models
(``vertex/edge updates can be modeled as tasks''), declarative programming (``graph operations can be written as datalog 
programs''), and domain-specific languages (``graph processing needs
its own language''). Of all these models, vertex programming has been
quite popular due to ease of use and the wide variety of different frameworks supporting it \cite{SIGMOD:2014}. 
While vertex programming is generally productive for writing graph programs, it lacks a strong mathematical model and is therefore difficult to analyze for program behavior 
or optimize for better backend performance. Matrix models, on the other hand, are based on a solid mathematical 
foundation i.e. graph traversal computations are modeled as operations on a semi-ring \cite{combblas}. CombBLAS \cite{combblas} is an extensible distributed-memory parallel graph library offering a set of linear algebra primitives specifically targeting graph analytics. While this
model is great for reasoning and performing optimizations, it is seen as hard to program. As shown in \cite{SIGMOD:2014}, some graph computations such as triangle counting are hard to express efficiently as a 
pure matrix operation, leading to long runtimes and increased memory consumption.

In the High Performance Computing world, sparse matrices are widely used in simulations and modeling of 
physical processes. Sparse matrix vector multiply (SPMV) is a key kernel used in operations such as linear solvers and eigensolvers. A variety of optimizations have been performed to improve SPMV performance on single 
and multiple nodes \cite{SPMV}. 
Existing matrix-based graph analytics operations achieve nowhere near the same performance as these optimized routines. 
Our goal is to achieve ``vertex programming 
productivity with HPC-like performance for graph analytics''.

There have been a large number of frameworks proposed for graph
analytics recently, and these differ both in terms of programming
abstractions as well as underlying implementations. There has been
recent work ~\cite{SIGMOD:2014} that has compared different graph
frameworks including Giraph~\cite{giraph} and GraphLab~\cite{graphlab,
Low:2010:UAI} which are two popular vertex programming models;
CombBLAS~\cite{combblas, Buluc:2011:HPCA}, a matrix programming
model; Socialite~\cite{Seo:2013:ICDE}, a functional programming
model; and Galois~\cite{Pingali:2011:PLDI, galois,
Nguyen:2013:SOSP}, a task-based abstraction. That paper shows that
CombBLAS and Galois generally perform well compared to other
frameworks. Moreover, the ability to map many diverse graph operations
to a small set of matrix operations means that the backend of CombBLAS
is easy to maintain and extend -- for example to multiple nodes
(Galois does not yet have a multi-node version). Hence, in terms of
performance, we can conclude that matrix-based abstractions are
clearly a good choice for graph analytics. Matrices are becoming an important class of objects in databases. Our technique of looking at graph algorithms as generalizations of sparse matrix algebra leads to a simple way to connect graph stores to array databases. We believe the rise of sparse array based databases will also help the use of graph storage and analytics. 

There are other matrix
based frameworks such as PEGASUS \cite{pegasus} for graph processing. PEGASUS is based on
Map-Reduce and suffers from poor performance due to I/O bottlenecks compared to in-memory frameworks. Other domain specific
languages such as GreenMarl \cite{greenmarl} purport to improve
productivity and performance, but at the cost of a having to learn a
new programming language.  Some other ways to process graphs include writing vertex programs as UDFs for use in a column store \cite{Alekh} and GraphX \cite{graphx} (set of graph primitives intended to work with Spark \cite{spark}). The popularity and adoption of vertex based
programming models (for instance, Facebook uses
Giraph~\cite{facebook:giraph}) establishes the case for vertex-based models over other alternatives.

In this work, we try to adopt the best of both worlds, and we
compare ourselves to high performing vertex programming and matrix
programming models (GraphLab and CombBLAS respectively). We will focus
on comparing GraphMat to GraphLab, CombBLAS and Galois for the reminder of this paper.

\section{Algorithms} \label{sec:algorithms}

To showcase the performance and productivity of GraphMat, 
we picked five different algorithms from a diverse set of applications,
including machine learning, graph traversal and graph statistics. Our choice covers a 
wide range of varying functionality (e.g. traversal or statistics), 
data per vertex, amount of communication,  iterative vs. non iterative etc. 
We give a brief summary on each algorithm below.

{\bf I. Page Rank (PR)}: This is an iterative algorithm used to rank web pages 
based on some metric (e.g. popularity). The idea is compute 
the probability that a random walk through the hyperlinks (edges)
would end in a particular page (vertex). The algorithm iteratively
updates the rank of each vertex according to the following equation: 

\begin{equation}
PR^{t+1}(v) = r + (1-r)* \sum_{u|(u,v) \in E} \frac{PR^{t}(u)}{\textrm {degree}(u)} \label{eq:pagerank}
\end{equation}

where $PR^t(v)$ denotes the page rank of vertex $v$ at iteration $t$, 
$E$ is the set of edges in a directed graph, and $r$ is the probability of random surfing. 
The initial ranks are set to $1.0$.  

\vspace{2mm}

{\bf II. Breadth First Search (BFS)}: This is a very popular graph search algorithm, 
which is also used as the kernel by the Graph500 benchmark \cite{graph500}.
The algorithm begins at a given vertex (called {\em root}) and iteratively
explores all connected vertices of an undirected and unweighted graph.     
The idea is to assign a distance to each vertex, where 
the distance represents the minimum number of edges needed to be
traversed to reach the vertex from the root. Initially, the distance of the root is set to $0$ and it is marked active. The other distances are set to infinity. At iteration $t$, each vertex adjacent to an active vertex computes the following:

\begin{equation}
Distance(v) = \min (Distance(v), t+1)
\end{equation}

If the update leads to a change in distance (from infinity to $t+1$), then the vertex becomes active for the next iteration. 

\vspace{2mm}

{\bf III. Collaborative Filtering (CF)}: This is a machine learning algorithm used
by many recommender systems \cite{ricci2011introduction} for 
estimating a user's rating for a given item based
on an incomplete set of (user, item) ratings. 
The underlying assumption is that users' ratings are based on a set of hidden/latent features and each item can be expressed as a combination of these features. Ratings depend on how well the user's and item's features match.
Given a matrix ${\bf G}$ of ratings, 
the goal of collaborative filtering technique 
is to compute two factors ${\bf P_U}$ and ${\bf P_V}$, 
each one is a low-dimensional dense matrix. This can be  
accomplished using incomplete matrix factorization \cite{matrixfactorization}. 
Mathematically, the problem can be expressed as eq. (\ref{eq:sgd}) where $u$ and $v$ are
the indices of the users and items, respectively, 
${\bf G}_{uv}$ is the rating of the $u^{th}$ user for the $v^{th}$ item, 
${\bf p}_u \& {\bf p}_v$ are dense vectors of length $K$ corresponding to 
each user and item, respectively. 

\begin{equation}
\min_{{\bf P_U,P_V}} \sum_{(u,v)\in G} ({\bf G}_{uv} - {\bf p}_u^T{\bf p}_v)^2 + \lambda ||{\bf p}_u||^2 + \lambda ||{\bf p}_v||^2 \label{eq:sgd}
\end{equation}

Matrix factorization is usually performed iteratively using Stochastic Gradient Descent (SGD) or Gradient Descent (GD). 
In each iteration $t$, GD performs Equation \ref{eq:sgd1} - \ref{eq:sgd3} for all users and items. SGD performs the same updates without the summation in equation \ref{eq:sgd2} on all ratings in a random order.  
The main difference between GD and SGD is that GD updates all the ${\bf p}_u$ and ${\bf p}_v$ once per iteration
instead of once per rating as in SGD.

\begin{eqnarray}
e_{uv} = {\bf G}_{uv} - {\bf p}_u^T{\bf p}_v \label{eq:sgd1} \\
{\bf p}_u^* = {\bf p}_u + \gamma [\sum_{(u,v)\in G} e_{uv}{\bf p}_v - \lambda {\bf p}_u] \label{eq:sgd2} \\
{\bf p}_v^* = {\bf p}_v + \gamma [\sum_{(u,v)\in G} e_{uv}{\bf p}_u - \lambda {\bf p}_v] \label{eq:sgd3} 
\end{eqnarray}

\vspace{2mm}
{\bf IV. Triangle Counting (TC)}: This is a statistics algorithm useful for understanding
social networks, graph analysis and computing clustering coefficient. The algorithm
computes the number of triangles in a given graph. A triangle exists when a vertex
has two adjacent vertices that are also adjacent to each other. The technique
used to compute the number of triangles is as follows. Each vertex shares its 
neighbor list with each of its neighbors. Each vertex then computes the intersection 
between its neighbor list and the neighbor list(s) it receives. For a given directed
graph with no cycles, the size of the intersections gives the number of 
triangles in the graph. When the graph is undirected, then each vertex in a triangle
contributes to the count, hence the size of the intersection is exactly 
3 times the number of triangles. The problem can be expressed
mathematically as follows, where $E_{uv}$ denotes the presence of an 
(undirected) edge between vertex $u$ and vertex $v$.

\begin{equation}
N_{triangles} = \sum_{u,v,w \in V| u<v<w} (u,v) \in E \land (v,w) \in E \land (u,w) \in E
\end{equation}

\vspace{2mm}
{\bf V. Single Source Shortest Path (SSSP)}: This is another graph algorithm used to
compute the shortest paths from a single source to all other vertices in a given weighted
and directed graph. The algorithm is used in many applications such as finding driving 
directions in maps or computing the min-delay path in telecommunication networks.    
Similar to BFS, the algorithm starts with a given vertex (called {\em source}) and
iteratively explores all the vertices in the graph. The idea is to assign a
distance value to each vertex, which is the minimum edge weights needed to reach 
a particular vertex from the source. At each iteration $t$,     
each vertex performs the following:

\begin{equation}
Distance(v) = \min_{u|(u,v) \in E} \{Distance(u)+ w(u,v)\}
\end{equation}

Where $w(u, v)$ represents the weight of the edge $(u, v)$. Initially the $Distance$ for each vertex
is set to infinity except the source with $Distance$ value set to 0. We use a slight variation on the Bellman-Ford shortest path algorithm where we only update the distance of those vertices that are adjacent to those that changed their distance in the previous iteration.

We now discuss the implementation of GraphMat and its optimizations in the next section.

\section{GraphMat} \label{sec:graphmat}

\begin{figure}[htpb!]
\begin{center}
\includegraphics[width=\columnwidth]{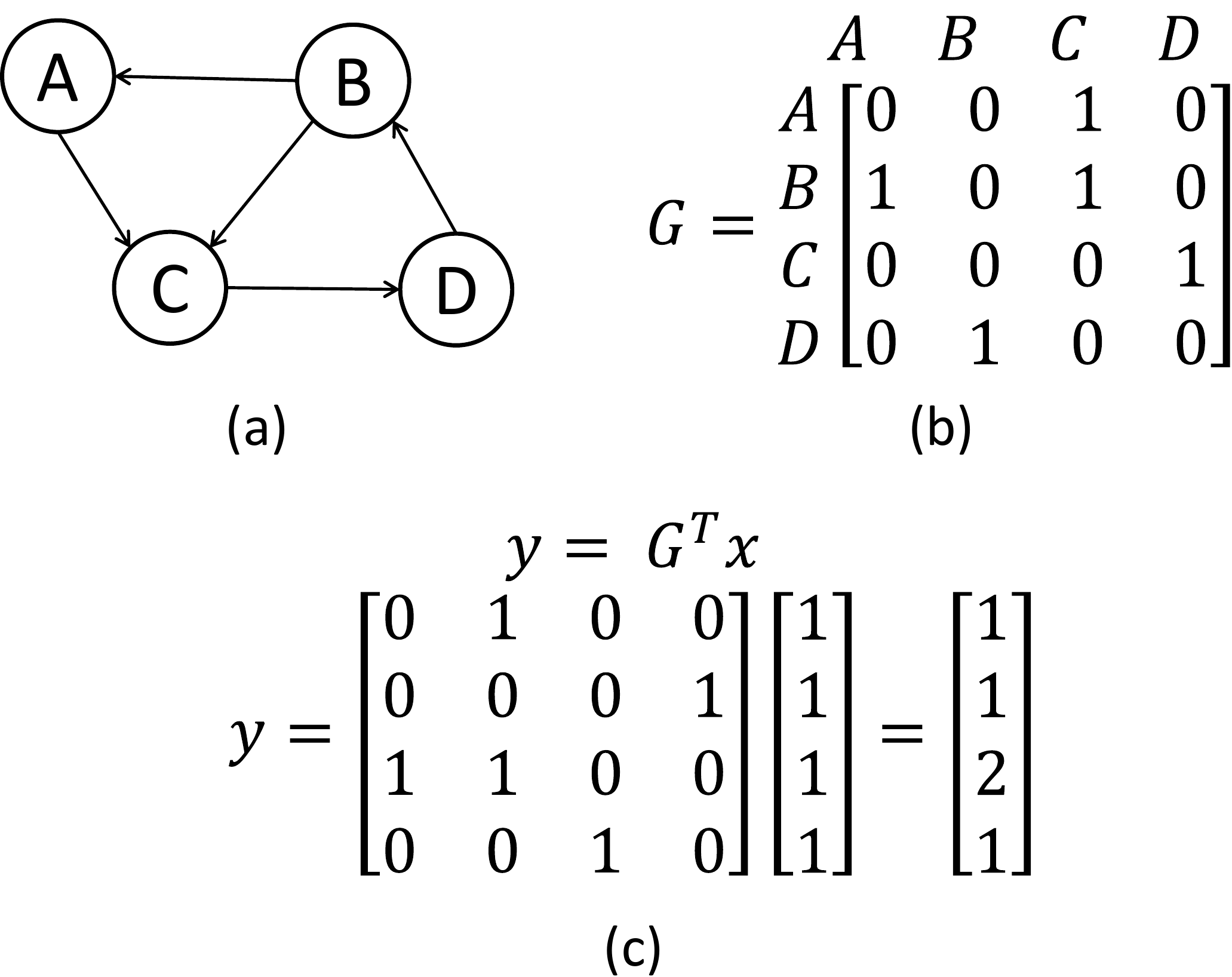}
\caption{Graph (a) Logical representation (b) Adjacency matrix (c) In-degree calculation as SPMV $G^Tx=y$. Vector $x$ is all ones. The output vector $y$ indicates the number of incoming edges for each vertex.}
\label{fig:vertex-vs-matrix}
\end{center}
\end{figure}

GraphMat is based on the idea that graph analytics via vertex programming can be performed through a backend that supports only sparse matrix operations. GraphMat takes graph algorithms written as vertex programs and performs generalized sparse matrix vector multiplication on them (iteratively in many cases). This is possible as edge traversals from a set of vertices can be written as sparse matrix-sparse vector multiplication routines on the graph adjacency matrix (or its transpose). To illustrate this idea, a simple example of calculating in-degree is shown in Figure \ref{fig:vertex-vs-matrix}. Multiplying the transpose of the graph adjacency matrix (unweighted graph) with a vector of all ones produces a vector of vertex in-degrees. To get the out-degrees, one can multiply the adjacency matrix with a vector of all ones.

\subsection{Mapping Vertex Programs to Generalized SPMV}
\label{subsec:graphmatmapping}
\begin{figure}[htpb!]
\begin{center}
\includegraphics[scale=0.35, angle=270]{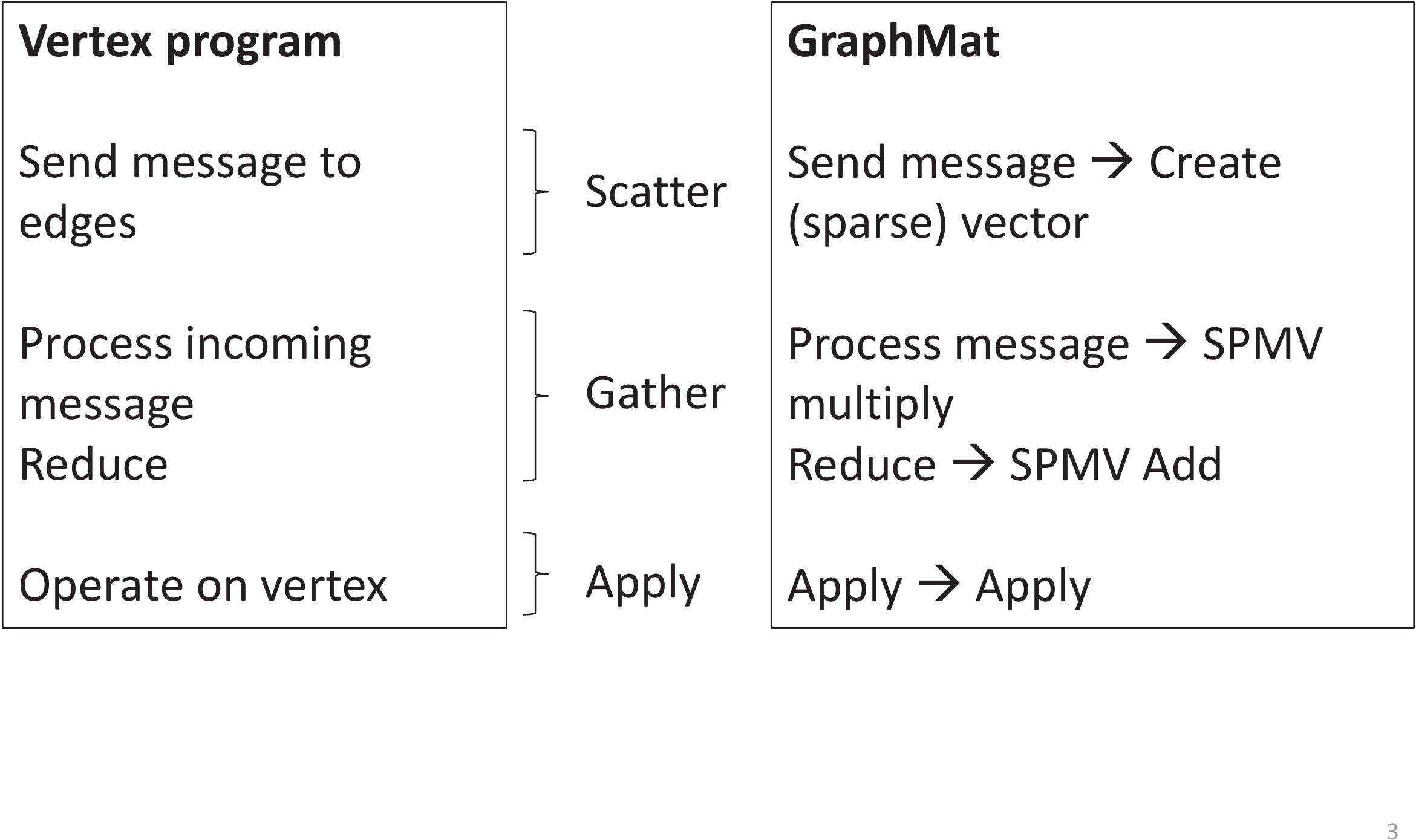}
\caption{Overview of vertex program to Sparse matrix vector multiply conversion.}
\label{fig:vertex-to-spmv-overview}
\end{center}
\end{figure}

The high-level scheme for converting vertex programs to sparse matrix programs is shown in Figure \ref{fig:vertex-to-spmv-overview}. We observe that while vertex programs can have slightly different semantics, they are all equivalent in terms of expressibility. Our vertex programming model is similar to that of Giraph \cite{giraph}. 

A typical vertex program has a state associated with each vertex that is updated iteratively. Each iteration starts with a subset of vertices that are ``active'' i.e. whose states were updated in the last iteration, which now have to broadcast their current state (or a function of their current state) to their neighboring vertices. A vertex receiving such ``messages'' from its neighbors processes each message separately and reduces them to a single value. The reduced value is used to update the current state of the vertex. Vertices that change state then become active for the next iteration. The iterative process continues for a fixed number of iterations or until no vertices change state (user-specified termination criterion). We follow the Bulk-synchronous parallel model i.e. each iteration can be considered a superstep.

The user specifies the following for a graph program in GraphMat - each vertex has user-defined property data that is initialized (based on the algorithm used). A set of vertices are marked active. The user-defined function {\sc Send\_Message}() reads the vertex data and produces a message object (done for each active vertex), {\sc Process\_Message}() reads the message object, edge data along which the message arrived, and the destination vertex data and produces a processed message for that edge. The {\sc Reduce}() function is typically a commutative function taking in the processed messages for a vertex and producing a single reduced value. {\sc Apply}() reads the reduced value and modifies its vertex data (done for each vertex that receives a message). {\sc Send\_Message}() can be called to scatter along in- and/or out- edges. We found that this model was sufficient to express a large number of diverse graph algorithms efficiently. The addition of access to the destination vertex data in {\sc Process\_Message}() makes algorithms like triangle counting and collaborative filtering easier to express than traditional matrix based frameworks such as CombBLAS. See Section \ref{section:genspmv} for more details.

\begin{figure*}[htpb!]
\begin{center}
\includegraphics[width=1.3\columnwidth]{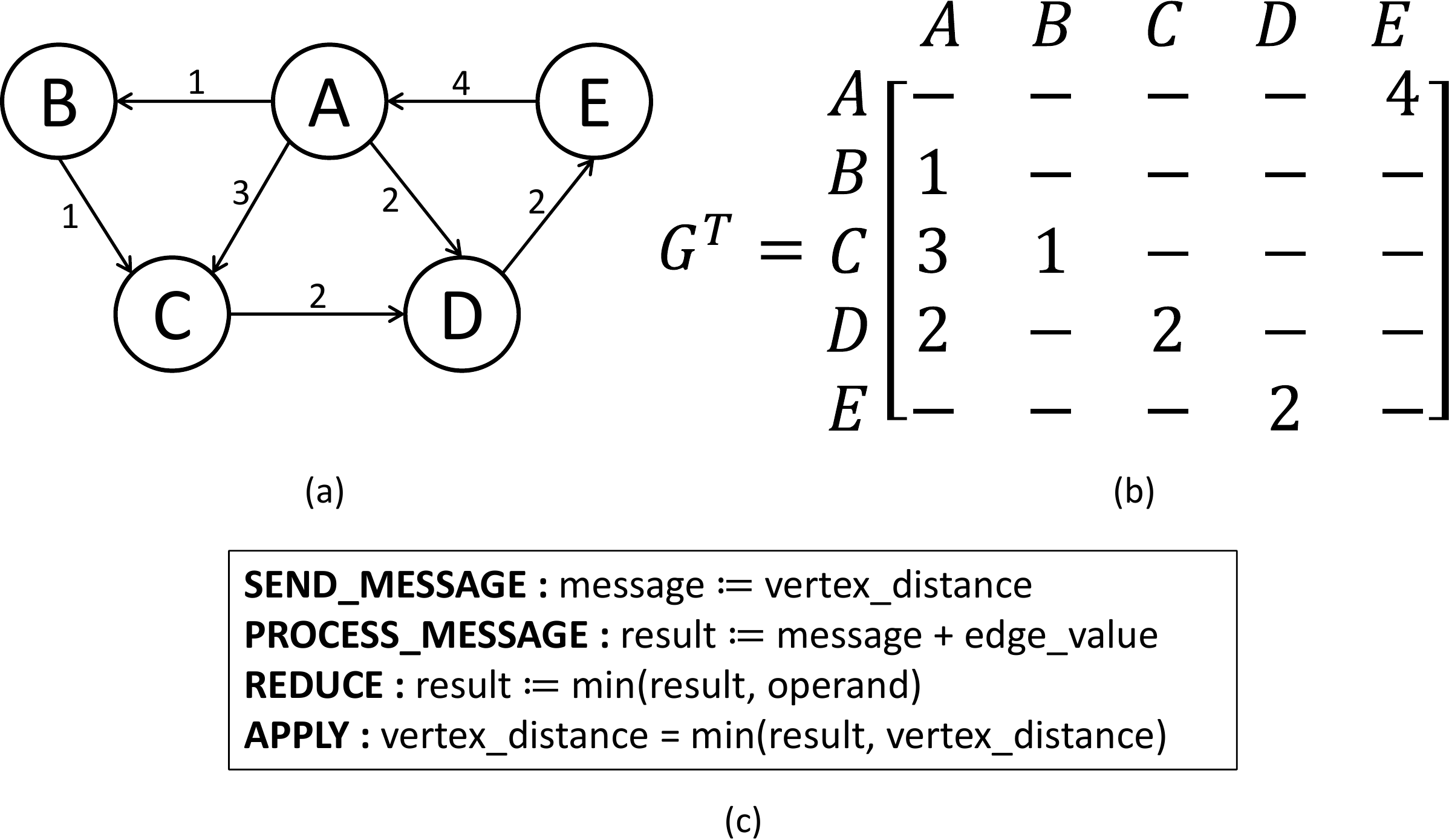}
\includegraphics[width=1.9\columnwidth]{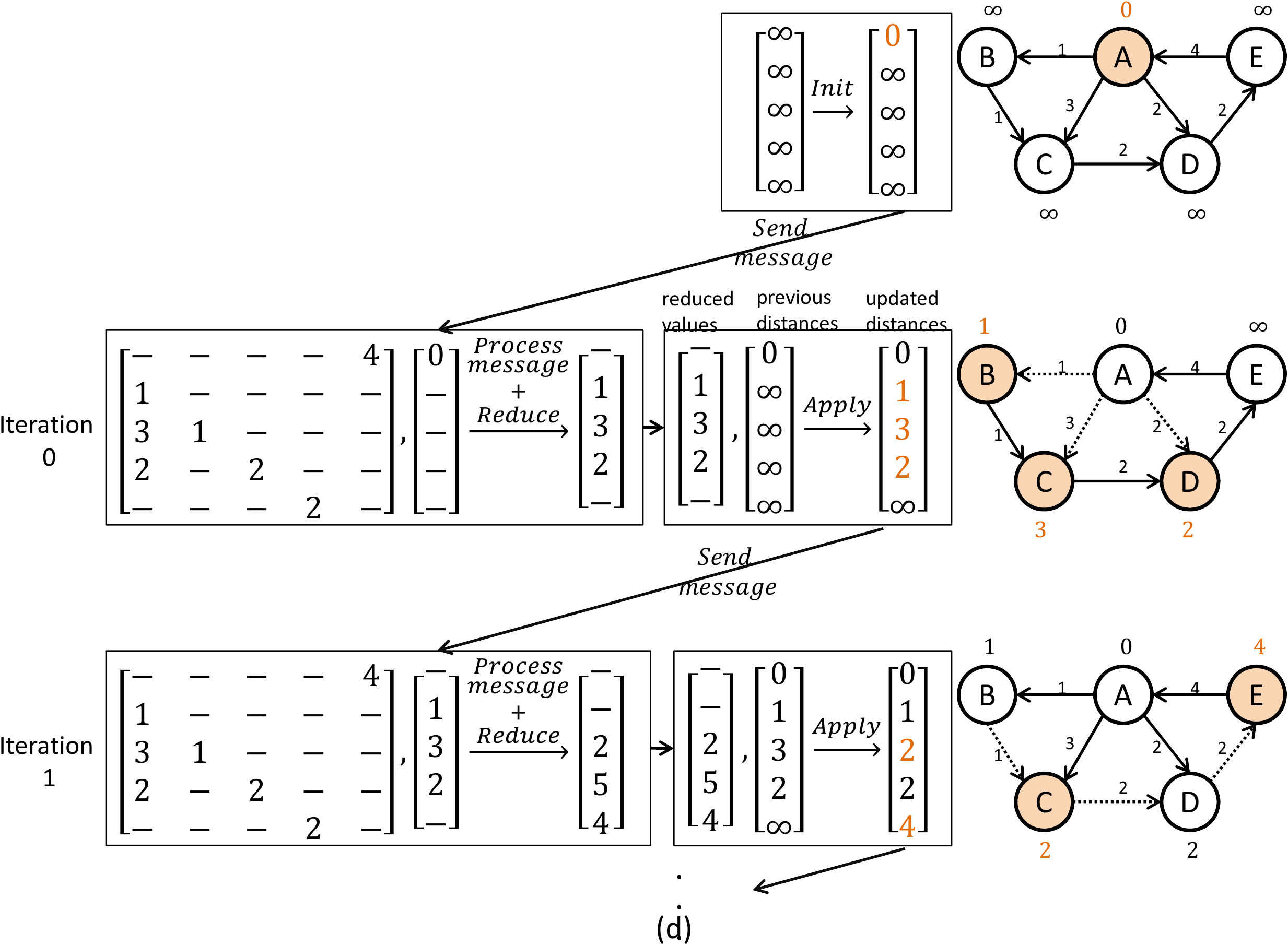}
\caption{Example: Single source shortest path. (a) Graph with weighted edges. (b) Transpose of adjacency matrix (c) Abstract GraphMat program to find the shortest distance from a source. (d) We find the shortest distance to every vertex from vertex A. Each iteration shows the matrix operation being performed ({\sc Process\_Message} and {\sc Reduce}). Dashed entries denote edges/messages that do not exist (not computed). The final vector (after {\sc Apply}) is the shortest distance calculated so far. On the right, we show the operations on the graph itself. Dotted lines show the edges that were processed in that iteration. Vertices that change state in that iteration and are hence active in the next iteration are shaded. The procedure ends when no vertex changes state. Figure best viewed in color.}
\label{fig:sssp-spmv}
\end{center}
\end{figure*}

Figure \ref{fig:sssp-spmv} shows an example of single source shortest path executed using the user-defined functions used in GraphMat. We calculate the shortest path to all vertices from source vertex A. At a given iteration, we generate a sparse vector using the {\sc Send\_Message}() function on the active vertices. The message is the shortest distance to that vertex calculated so far. {\sc Process\_Message}() adds this message to the edge length, while {\sc Reduce}() performs a min operation. {\sc Process\_Message}() and {\sc Reduce}() together form a sparse matrix sparse vector multiply operation replacing traditional SPMV multiply operation with addition and SPMV addition with min respectively. Source code for the SSSP algorithm in GraphMat is provided in the appendix.

\subsection{Generalized SPMV} \label{section:genspmv}

As shown in Figures \ref{fig:vertex-vs-matrix} and \ref{fig:sssp-spmv}, a generalized sparse matrix vector multiplication helps implement multiple graph algorithms. These examples, though simple, illustrate that overloading the multiply and add operations of a SPMV can produce different graph algorithms. In this framework, a vertex program with {\sc Process\_Message} and {\sc Reduce} functions can be written as a generalized SPMV. Assuming that the graph adjacency matrix transpose $G^T$ is 
stored in a Compressed Sparse Column (CSC) format, a generalized SPMV is given in Algorithm \ref{alg:spmv}. We can also partition this matrix into many chunks to improve parallelism and load balancing.

\begin{algorithm}[!hbt]
{\small
\begin{spacing}{\algospacing} 
\caption{Generalized SPMV}
\label{alg:spmv}
\begin{algorithmic}[1]
\Function{SPMV}{Graph $G$, SparseVector ${\bf x}$, {\sc Process\_Message}, {\sc Reduce}}
\State ${\bf y} \leftarrow$ new SparseVector()
\For{$j$ in $G^T$.column\_indices}
	\If{$j$ is present in ${\bf x}$}
		\For{$k$ in $G^T$.column$_j$}
			\State result $\leftarrow$ {\sc Process\_Message}(${\bf x}_j$, $G$.edge\_value($k,j$),  $G$.getVertexProperty($k$))
			\State ${\bf y}_k$ $\leftarrow$ {\sc Reduce}(${\bf y}_k$, result)
		\EndFor
	\EndIf
\EndFor
\Return ${\bf y}$
\EndFunction
\end{algorithmic}
\end{spacing}
}
\end{algorithm}

We implement SPMV by traversing the non-zero columns in $G^T$. If a particular column $j$ has a corresponding non-zero at position $j$ in the sparse vector, then the elements in the column are processed and values accumulated in the output vector $y$.

GraphMat's main advantage over other matrix based frameworks is that it is easy for the user to write different graph programs with a vertex program abstraction. With other matrix-based frameworks such as CombBLAS\cite{combblas} and PEGASUS \cite{pegasus}, the user defined function to process a message (equivalent to GraphMat's {\sc Process\_Message}) can only access the message itself and the value of the edge along which it is received (similar to the example in Figure \ref{fig:vertex-vs-matrix}). This is very restrictive for many algorithms esp. Collaborative filtering and Triangle counting. In GraphMat, the message processing function can access the property data of the vertex receiving the message in addition to the message and edge value. We have found that this makes it very easy to write different graph algorithms with GraphMat. While one could technically achieve vertex data access during message processing with CombBLAS, it involves non-trivial accesses to the internal data structures that CombBLAS maintains, adding to coding complexity of pure matrix based abstractions. For example with triangle counting, a straightforward
implementation in CombBLAS uses a matrix-matrix multiply which
results in long runtimes and high memory
consumption~\cite{SIGMOD:2014}. 
Triangle Counting in GraphMat works as two vertex programs. The first 
creates an adjacency list of the graph (this is a simple vertex
program where each vertex sends out its id, and at the end stores a list of all its
incoming neighbor id's in its local state). In the second program,
each vertex simply sends out this list to all neighbors, and each vertex
intersects each incoming list with its own list to find triangles
(as described in Section~\ref{sec:algorithms}-IV). This approach is more efficient and is faster. 
Similar issues occur with implementing
Collaborative Filtering in CombBLAS as well. 

\subsection{Overall framework}

The overall GraphMat framework is presented in Algorithm \ref{alg:graphmat}. The set of active vertices is maintained using a boolean array for performance reasons. In each iteration, this array is scanned to find the active vertices and a sparse vector of messages is generated. Then, a generalized SPMV is performed using this vector. The resulting output vector is used to update the state of the vertices. If any vertices change state, they are marked active for the next iteration. The algorithm continues for a user-specified maximum number of iterations or until convergence (no vertices change state).

\begin{algorithm}[!hbt]
{\small
\begin{spacing}{\algospacing} 
\caption{GraphMat overview. ${\bf x, y}$ are sparse vectors.}
\label{alg:graphmat}
\begin{algorithmic}[1]
\Function{Run\_Graph\_Program}{Graph $G$, GraphProgram $P$}
\For{$i=1$ to $MaxIterations$}
	\For{$v=1$ to $Vertices$}
		\If{$v$ is $active$}
			\State ${\bf x}_v$ $\leftarrow$ {\sc $P$.Send\_Message}($v$, $G$)
		\EndIf
	\EndFor
	\State ${\bf y}$ $\leftarrow$ SPMV($G$, ${\bf x}$, {\sc $P$.Process\_Message}, {\sc $P$.Reduce})
	\State Reset $active$ for all vertices
	\For{$j=1$ to ${\bf y}$.length}
		\State $v$ $\leftarrow$ ${\bf y}$.getVertex($j$)
		\State $old\_vertexproperty$ $\leftarrow$ $G$.getVertexProperty($v$)
		\State $G$.setVertexProperty($v$, ${\bf y}$.getValue($j$), {\sc $P$.Apply})
		\If{$G$.getVertexProperty($v$) $\neq$ $old\_vertexproperty$}
			\State $v$ set to $active$
		\EndIf
	\EndFor
	\If{Number of $active$ vertices == 0}
		\State break
	\EndIf	
\EndFor

\EndFunction
\end{algorithmic}
\end{spacing}
}
\end{algorithm}

As shown in Algorithm \ref{alg:graphmat}, GraphMat follows an iterative process of {\sc Send\_Message} (lines 3-5), SPMV (line 6), and {\sc Apply} (lines 8-13). Each such iteration is a superstep.

\subsection{Data structures}

We describe the sparse matrix and sparse vector data structures in this section. 

\subsubsection{Sparse Matrix}
We represent the sparse matrix in the Doubly Compressed Sparse Column (DCSC) format~\cite{BulucG08} which can store very large sparse matrices efficiently.
It primarily uses four arrays to store a given matrix as briefly explained here:
one array to store the column indices of the columns which have at-least one non-zero element,
two arrays storing the row indices (where there are non-zero elements) corresponding to each of the above column indices and the non-zero values themselves, and
another array to point where the row-indices corresponding to a given column index begin in the above array (allowing access to any non-zero element at a given column index and a row index if it is present).
The format also allows an optional array to index the column indices with non-zero elements, which we have not used.
For more details and examples, please see~\cite{BulucG08}.
The DCSC format has been used effectively in parallel algorithms for problems such as Generalized sparse matrix-matrix multiplication (SpGEMM)~\cite{BulucG12}, and is part of the Combinatorial BLAS (CombBLAS) library~\cite{Buluc:2011:HPCA}. The matrix is partitioned in a 1-D fashion (along rows), and each partition is stored as an independent DCSC structure.

\subsubsection{Sparse Vector}
Sparse Vectors can be implemented in many ways. Two good ways of storing sparse vectors are as follows: (1) A variable sized array of sorted (index, value) tuples (2) A bitvector for storing valid indices and a constant (number of vertices) sized array with values stored only at the valid indices. Of these, the latter option provides better performance across all algorithms and graphs and so is the only option considered for the rest of the paper. In the SPMV routine in Algorithm \ref{alg:spmv}, line 4 becomes faster due to use of the bitvector. Since the bitvector can also be shared among multiple threads and can be cached effectively, it also helps in improving parallel scalability. 
The performance gain from this bitvector use is presented in Section \ref{sec:results}. 

\vspace{0.5in}

\subsection{Optimizations}
\label{subsec:graphmatopt}
Some of the optimizations performed to improve the performance of GraphMat are described in this section. The most important optimizations improve the performance of the SPMV routine as it accounts for most of the runtime. 
\begin{enumerate}
\item Cache optimizations such as the use of bitvectors for storing sparse vectors improve performance.
\item Since the generalized SPMV operations ({\sc Process\_Message} and {\sc Reduce}) are user-defined, using the compiler option to perform inter-procedural optimizations (-ipo) is essential.
\item Parallelization of SPMV among multiple cores in the system increases processing speed. Each partition of the matrix is processed by a different thread.
\item Load balancing among threads can be improved through better partitioning of the adjacency matrix. We partition the matrix into many more partitions than number of threads along with dynamic scheduling to distribute the SPMV load among threads better. Without this load balancing, the number of graph partitions equals number of threads.
\end{enumerate}

We now discuss the experimental setup, datasets used and the results of our comparison to other graph frameworks.

\section{Results} \label{sec:results}

\subsection{Experimental setup}
\label{subsec:exptsetup}

We performed the experiments \footnote{\scriptsize Software and workloads used in performance tests may have been optimized for performance only on Intel microprocessors.  Performance tests, such as SYSmark and MobileMark, are measured using specific computer systems, components, software, operations and functions.  Any change to any of those factors may cause the results to vary.  You should consult other information and performance tests to assist you in fully evaluating your contemplated purchases, including the performance of that product when combined with other products.  
For more information go to http://www.intel.com/performance
} 
on an {Intel$^{\mbox{\tiny\textregistered}}$ } {Xeon$^{\mbox{\tiny\textregistered}}$ } 
\footnote{\scriptsize Intel and Xeon are trademarks of Intel Corporation in the U.S. and/or other countries.} 
E5-2697 v2 based system. The system contains two processors, each with 12 cores running at 2.7GHz (24 cores in total)
sharing 30 MB L3 cache and 64 GB of memory. The machine runs   
Red Hat Enterprise Linux Server OS release 6.5. 
We used the 
{Intel$^{\mbox{\tiny\textregistered}}$ } C++ Composer XE 2013 SP1 Compiler
\footnote{\scriptsize Intel's compilers may or may not optimize to the same degree for non-Intel
microprocessors for optimizations that are not unique to Intel microprocessors.
These optimizations include SSE2, SSE3, and SSE3 instruction sets and
other optimizations. Intel does not guarantee the availability, functionality,
or effectiveness of any optimization on microprocessors not manufactured by
Intel. Microprocessor-dependent optimizations in this product are intended
for use with Intel microprocessors. Certain optimizations not specific to Intel
micro-architecture are reserved for Intel microprocessors. Please refer to the
applicable product User and Reference Guides for more information regarding
the specific instruction sets covered by this notice. Notice revision \#20110804} 
and the {Intel$^{\mbox{\tiny\textregistered}}$ } MPI library 5.0 to compile 
the native and benchmark code.
We used GraphLab v2.2~\cite{graphlab}, CombBLAS v1.3 ~\cite{combblas} and Galois v2.2.0 ~\cite{galois} for performance comparisons. 
In order to utilize multiple threads on the CPU, GraphMat and Galois use OpenMP only,       
GraphLab uses both OpenMP and MPI, and CombBLAS uses MPI only. 
Since CombBLAS requires the total number of 
processes to be a square (due to their 2D partitioning
approach), we use 16 MPI processes to run on the 24 cores system (hence 8 cores
remain idle). We found that running CombBLAS with 25 MPI processes using
24 cores yields worse performance than running with 16 processes. However the native code, GraphMat, Galois, and GraphLab use the entire system (24 cores).

\textbf{Datasets:} 
We used a mix of real-world and synthetic datasets for our evaluations. 
Real-world datasets include Facebook interaction graphs~\cite{facebook09}, the Netflix challenge for collaborative filtering~\cite{netflix}, USA road network for California and Nevada~\cite{dimacs} and the Livejournal, Wikipedia, Delaunay and Flickr graphs from the University of Florida Sparse Matrix collection~\cite{Florida10}. 
We chose these datasets primarily to match those used in previous work~\cite{SIGMOD:2014,sssp-gpu-ipdps-2014,Nguyen:2013:SOSP} so that valid performance comparisons can be made.
Table~\ref{tab:realworld} provides details on the datasets used, as well as the algorithms run on these graphs. 

Since many real-world datasets are small in size, we augmented them
with synthetic datasets obtained from the Graph500 RMAT data
generator~\cite{graph500}. We adjust the RMAT parameters A,B,C,D
depending on the algorithm run (to correspond to previous work).
Specifically, following~\cite{SIGMOD:2014}, we use RMAT parameters A =
0.57, B=C= 0.19 (D is always = 1-A-B-C) for generating graphs for
Pagerank, BFS and SSSP; and different parameters A = 0.45, B=C =0.15
for Triangle Counting as in~\cite{SIGMOD:2014}. We generate one additional scale 24 graph for SSSP with parameters A=0.50, B=C=0.10 to match with that used in ~\cite{sssp-gpu-ipdps-2014,Nguyen:2013:SOSP}. Finally, for collaborative filtering, we used the synthetic bipartite graph generator as described in~\cite{SIGMOD:2014} to generate graphs similar in distribution to the real-world Netflix challenge graph.

Both real-world and synthetic graphs obtained occasionally need pre-processing for specific algorithms. We first remove self-loops in the graphs. Pagerank and SSSP usually assume all edges in the graph are directed and work directly with the graphs obtained. For BFS, we replicate edges (if the original graph is directed) to obtain a symmetric graph. For Triangle Counting, the input graph is expected to be directed acyclic; hence we first replicate edges as in BFS to make the graph symmetric and then discard the edges in the lower triangle of the adjacency matrix. Finally, for collaborative filtering, the graphs have to be bipartite; both the Netflix graph and synthetic graph generators ensure this. By using the same input graphs as previous work, we ensure that we can make direct performance comparisons.

\begin{table}[htbp]
  \centering
\resizebox{\columnwidth}{!} {
  \begin{tabular}{|c|c|c|c|c|}
   \hline
Dataset & \# Vertices & \# Edges & Algorithms & Brief Description \\
\hline
Synthetic                 & 1,048,576   & 16,746,179    & Tri Count,& Described in\\
Graph500~\cite{graph500}  &    &               & & Section ~\ref{subsec:exptsetup}         \\
RMAT Scale 20             &    &                 &         &          \\
\hline
Synthetic                 & 8,388,608   & 134,215,380   & Pagerank, BFS, & Described in\\
Graph500~\cite{graph500}  &    &               & SSSP & Section ~\ref{subsec:exptsetup}    \\
RMAT Scale 23             &    &                 &         &          \\
\hline
Synthetic                 & 16,777,216  & 267,167,794   & SSSP           & Described in\\
Graph500~\cite{graph500}  &    &               &                & Section ~\ref{subsec:exptsetup}    \\
RMAT Scale 24             &    &                 &         &          \\
\hline
LiveJournal~\cite{Florida10} & 4,847,571  & 68,993,773  & Pagerank, BFS, & LiveJournal \\
                             &            &             & Tri Count & follower graph\\
\hline
Facebook~\cite{facebook09}   & 2,937,612  & 41,919,708  & Pagerank, BFS, & Facebook user\\
                             &            &             & Tri Count & interaction graph\\
\hline
Wikipedia~\cite{Florida10}   & 3,566,908  & 84,751,827  & Pagerank, BFS,  & Wikipedia \\
                             &            &             & Tri Count & Link graph \\
\hline
Netflix~\cite{netflix}    & 480,189 users & 99,072,112  & Collaborative  & Netflix Prize\\
                          & 17,770 movies & ratings     & Filtering      & \\
\hline
Synthetic              & 63,367,472 users & 16,742,847,256   & Collaborative & Described in\\
Collaborative          & 1,342,176 items  & ratings          & Filtering     & Section ~\ref{subsec:exptsetup}\\
Filtering~\cite{SIGMOD:2014}              &                  &                  &               & \\
\hline
Flickr~\cite{Florida10}        & 820,878    & 9,837,214        & SSSP          & Flickr crawl\\
\hline
USA road ~\cite{dimacs}   & 1,890,815  & 4,657,742        & SSSP          & DIMACS9 \\
(CAL)                   &    &                 &         &          \\
\hline
  \end{tabular}
}
\caption{Real World and synthetic datasets}
  \label{tab:realworld}
\vspace*{-0.1in}
\end{table}

\subsection{Performance Results}

We first compare the runtime performance of GraphMat to other frameworks. We
demonstrate the performance improvement of GraphMat over a
common vertex programming framework (GraphLab \cite{graphlab}), a
high performance matrix programming framework (CombBLAS \cite{combblas}) and a high performance task based framework (Galois \cite{galois}). We then
compare GraphMat performance to that of native well-optimized hand-coded implementations of these
algorithms that gets performance limited only by
hardware~\cite{SIGMOD:2014}. Finally, we show the scalability of
GraphMat as compared to GraphLab, CombBLAS and Galois. 

\subsubsection{GraphMat vs. Other frameworks}

\begin{figure*}[htb]%
\begin{center}
        \subfigure[{PageRank}]
        {
                \label{fig:prsingle}
                \includegraphics[width=0.3\textwidth]{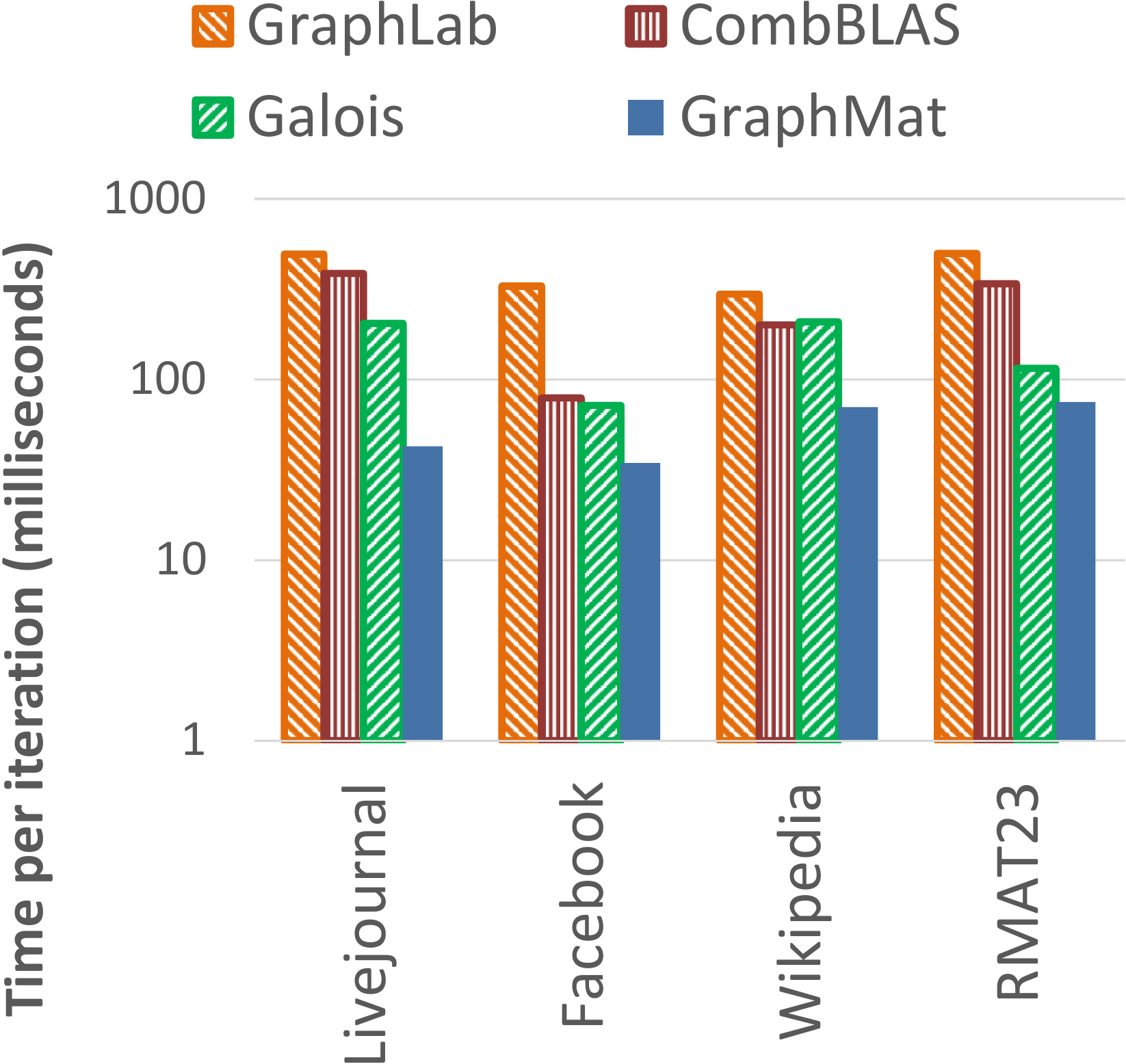}
        }
        \subfigure[{Breadth-First Search}]
        {
                \label{fig:bfssingle}
                \includegraphics[width=0.3\textwidth]{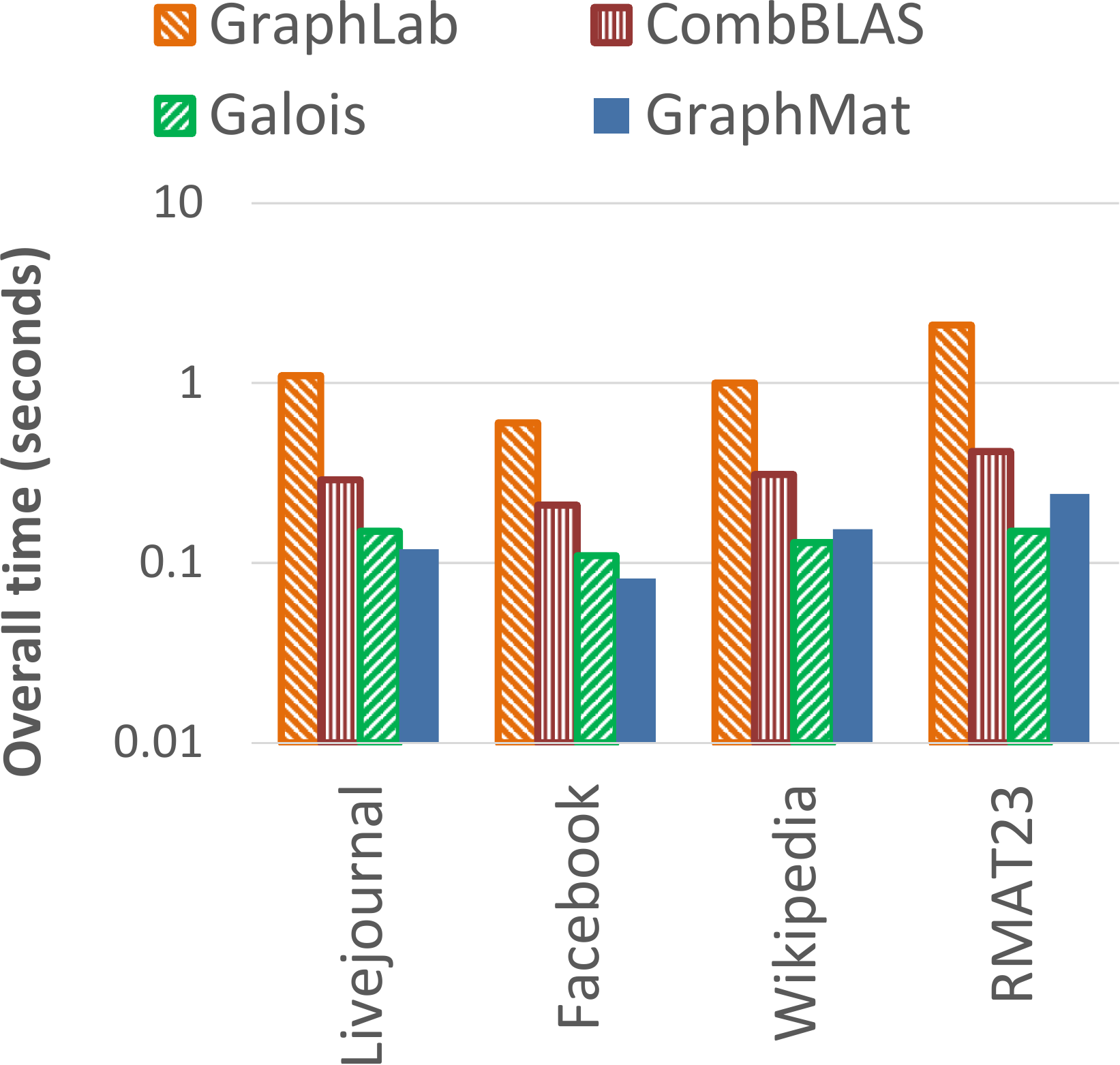}
        }
        \subfigure[{Triangle Counting}]
        {
                \label{fig:tcsingle}
                \includegraphics[width=0.3\textwidth]{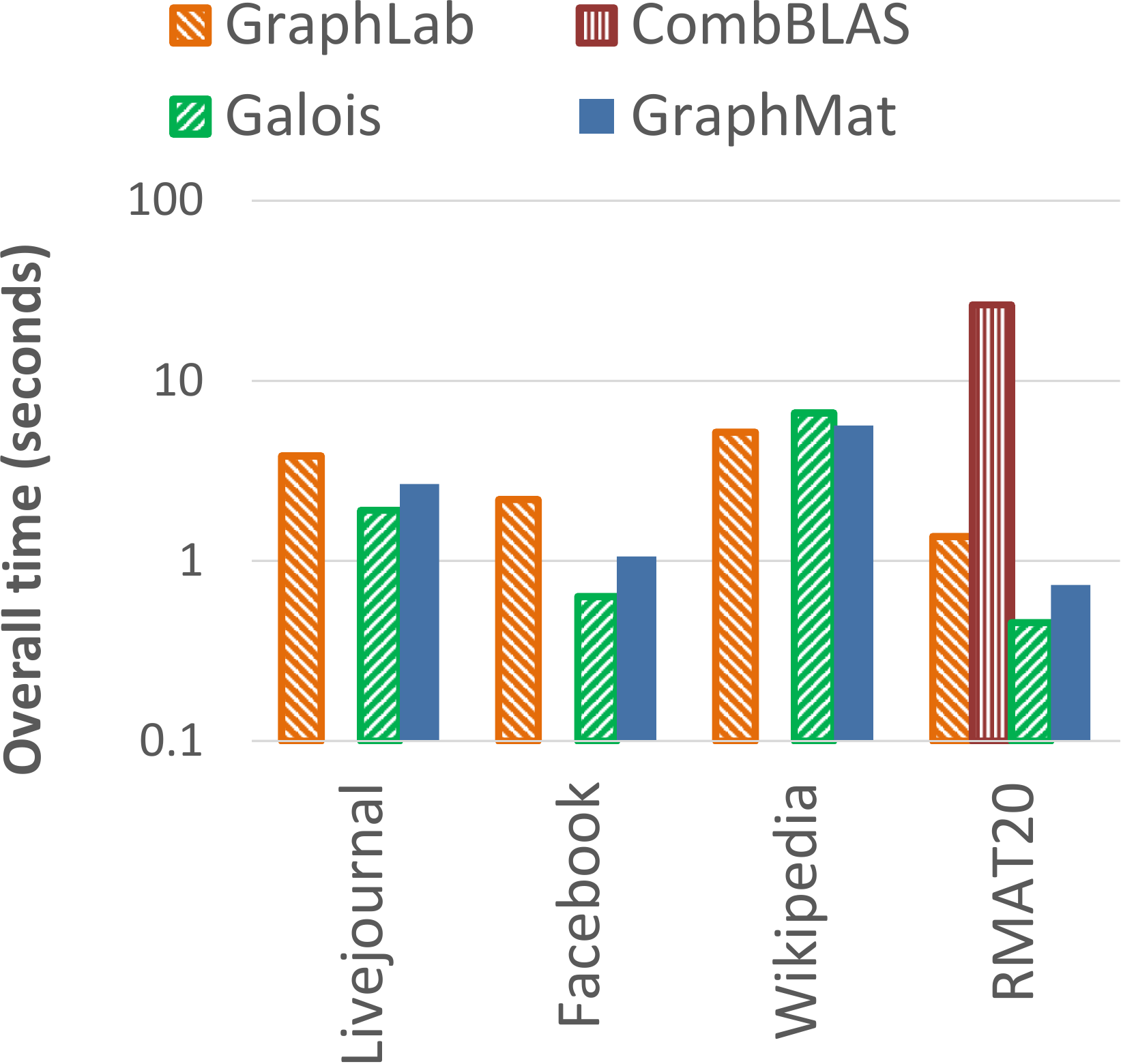}
        }
\\
        \subfigure[{Collaborative Filtering}]
        {
                \label{fig:cfsingle}
                \includegraphics[width=0.3\textwidth]{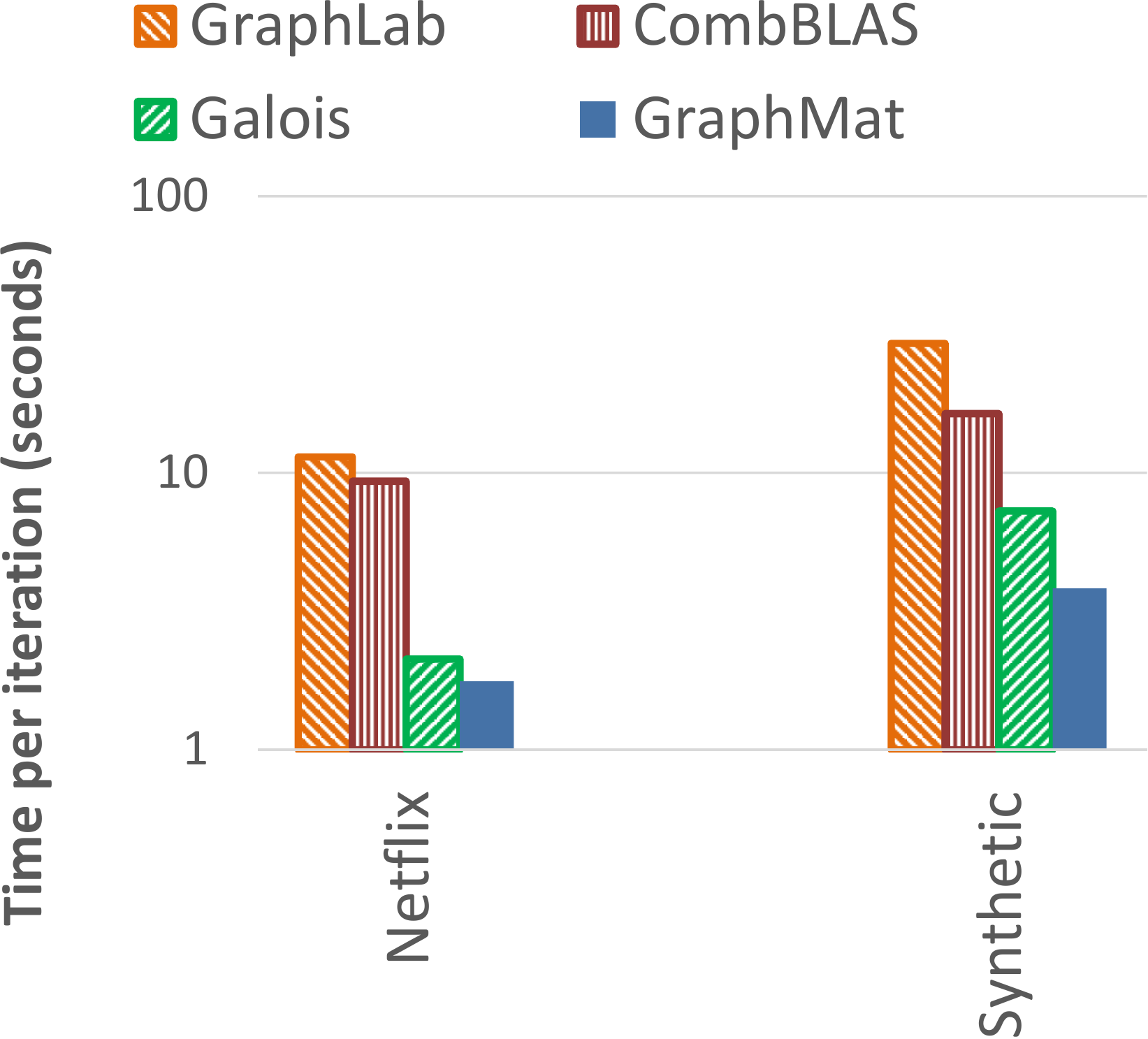}
        }
        \subfigure[{Single Source Shortest Path}]
        {
                \label{fig:ssspsingle}
                \includegraphics[width=0.3\textwidth]{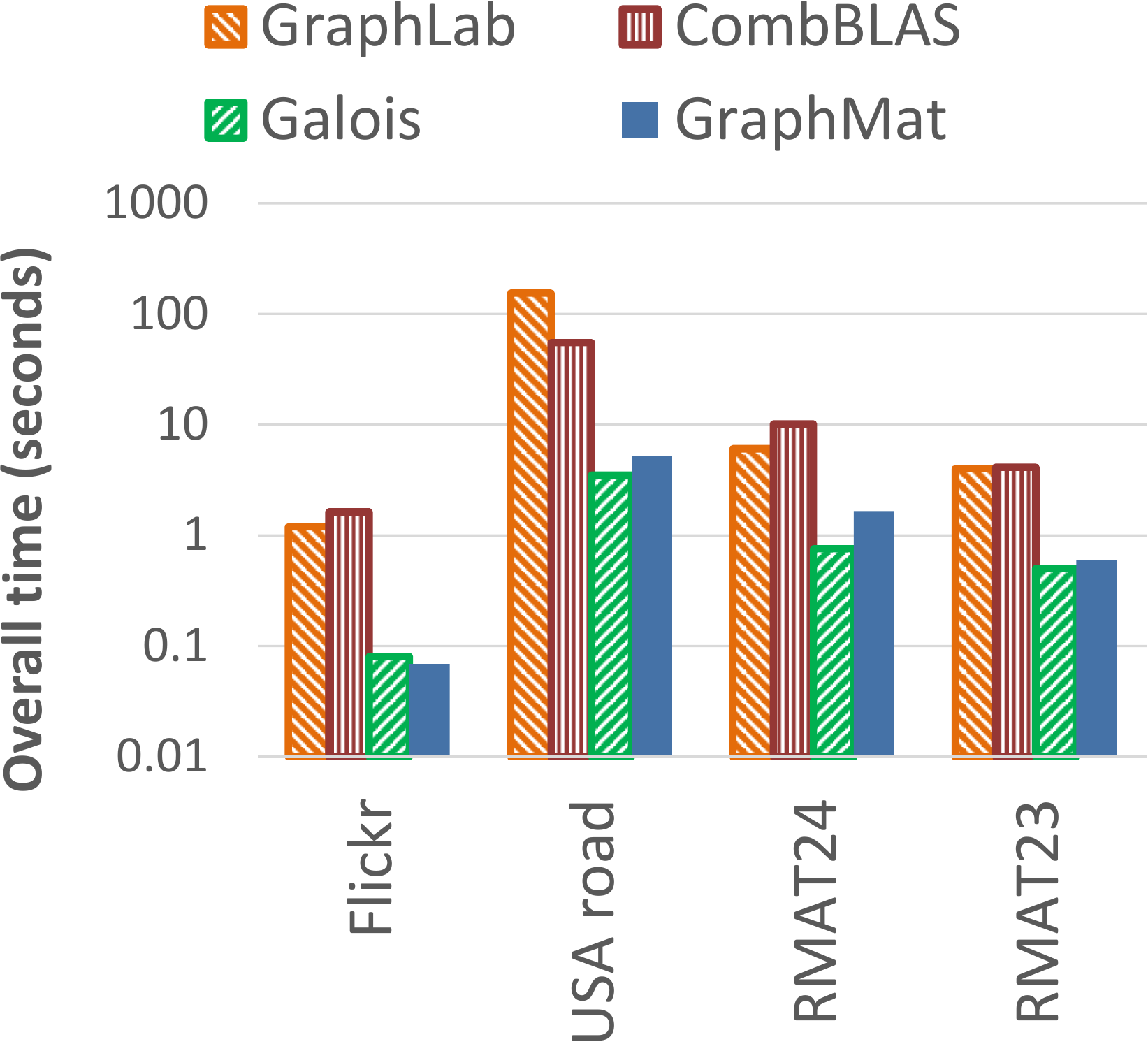}
        }
        \caption{\small Performance results for different algorithms
        on real-world and synthetic graphs. The y-axis represents runtime (in $\log$-scale), therefore lower numbers are better.}
        \label{fig:single}
\end{center}
\vspace*{-0.20in}
\end{figure*}

As mentioned in Section~\ref{sec:algorithms}, we selected a diverse set of graph algorithms, 
and used different real-world and synthetic datasets for these algorithms (see column ``Algorithms" 
in Table \ref{tab:realworld} for details) that were selected to be comparable to previous work. We report the 
time taken to run the graph algorithms after loading the graph into memory (excluding time taken to read the 
graph from disk).
Figure~\ref{fig:single} shows the performance results of running
GraphMat, GraphLab, CombBLAS and Galois on these algorithms and datasets. The
y-axis on these figures are total runtime, except for Pagerank and
Collaborative Filtering where each algorithm iteration takes similar
time and hence we report time/iteration. Since we report runtimes, lower bars indicate better performance.

We note that GraphMat is significantly faster than both GraphLab and
CombBLAS on most algorithms and datasets. GraphMat is faster than Galois on average.
As we can see from Figures~\ref{fig:prsingle} and ~\ref{fig:bfssingle}, GraphMat is 4-11X
faster than GraphLab on both real-world and synthetic datasets for
Pagerank and BFS (average of 7.5X for Pagerank and 7.9X for BFS).
As has been shown previously~\cite{SIGMOD:2014} on these datasets,
CombBLAS performs better than GraphLab due to its better optimized
backend, but GraphMat is still 2-4X better than CombBLAS. Compared to Galois, GraphMat is 1.5-4X better on Pagerank and ties on BFS. CombBLAS
performs poorly in Triangle Counting (Figure~\ref{fig:tcsingle}), where intermediate results 
are so large as to overflow memory or come close to memory limits;
CombBLAS fails to complete for real-world datasets and is about 36X
slower than GraphMat on the synthetic graph. GraphLab is much better
optimized for this algorithm due to the use of cuckoo hash data
structures and is only 1.5X slower than GraphMat on average. Galois is 20\% faster than GraphMat for triangle counting.
On Collaborative Filtering, Figure~\ref{fig:cfsingle} shows that
GraphMat is about 7X faster than GraphLab, 4.7X faster than
CombBLAS and 1.5X faster than Galois. These four algorithms were also studied
in~\cite{SIGMOD:2014}, and our performance results for GraphLab,
CombBLAS and Galois closely match the results in that paper. 

We consider an additional algorithm in this paper (SSSP) to increase the
diversity of applications covered. For Single Source Shortest Path (SSSP), GraphMat is about 10X faster
than both GraphLab and CombBLAS (Figure~\ref{fig:ssspsingle}). This difference is larger than ones
seen in other algorithms. This arises in part because some of these
datasets are such that SSSP takes a lot of iterations to finish with
each iteration doing a relatively small amount of work (especially for
Flickr and USA-Road graphs). For such computations, GraphMat, which
has a small per-iteration overhead performs much better than other
frameworks. For the other datasets that do more work per iteration,
GraphMat is still 3.6-6.9X better than GraphLab and CombBLAS. Galois 
performs better than GraphMat on SSSP by 30\%.

Table~\ref{tab:graphmat-graphlab-combBLAS} summarizes these results.
We see from the table that the geometric mean of the 
speedup of GraphMat over GraphLab and CombBLAS is about 5-7X and speedup over Galois is about 1.2X over the range of
algorithms and datasets. We defer a discussion of the
reasons for this performance difference to Section~\ref{subsec:perfdiscussion}. In the next section, we describe
how this performance compares to that of hand-optimized code, and then
discuss scalability of GraphMat. 

\begin{table}[htb!]
  \centering
{\small
  \begin{tabular}{|@{\hspace{3pt}}c@{\hspace{3pt}}|@{\hspace{3pt}}ccccc@{\hspace{3pt}}|@{\hspace{3pt}}c@{\hspace{3pt}}|}
   \hline

&PR  &      BFS  &   TC &       CF  &    SSSP  &   Overall \\
\hline

GraphLab & 	7.5	& 7.9	& 1.5	& 7.1	& 10.6	& 5.8 \\
CombBLAS &	4.1	& 2.2	&36.0	&4.8 	&10.2	& 6.9 \\
Galois & 2.6	& 1.0	& 0.8 &	1.5 & 0.7 & 1.2  \\

\hline
  \end{tabular}
}
\caption{Summary of performance improvement of
GraphMat over GraphLab, CombBLAS and Galois. Higher values mean GraphMat is faster. }
  \label{tab:graphmat-graphlab-combBLAS}
  \end{table}

\vspace{0.2 in}
 
\subsubsection{GraphMat vs. Native} 
We now compare GraphMat performance to that of hand-optimized native implementations. We took the 
performance results of native PageRank, BFS, Triangle counting, and
collaborative filtering implementations from \cite{SIGMOD:2014}, since
we used the same datasets and machines with identical configuration to
that work. %
Table~\ref{tab:native-vs-graphmat} shows the results of our comparison
with the geometric mean over all datasets for each algorithm. The table shows
the slowdown of GraphMat with
respect to native code. We can see that GraphMat is comparable in
performance for Pagerank and BFS. For Collaborative
Filtering, GraphMat is faster than native code in terms of runtime per iteration. This is because the native performance results from ~\cite{SIGMOD:2014} are
for Stochastic Gradient Descent (SGD) as opposed to Gradient Descent
(GD) for GraphMat, and GD is more easily parallelizable than SGD. This
is reflected in our performance results. 

Table~\ref{tab:native-vs-graphmat} shows that, on average, GraphMat is
only 1.2X slower than native code. It should be noted that
hand-optimized native code typically requires significant effort to
write even for expert users. Moreover, the effort is not usually very portable across
algorithms, and very specific tuning has to be done for each
algorithm and machine architecture. The efforts described in~\cite{SIGMOD:2014} are indeed
difficult to perform for an end-user of a graph framework. However,
GraphMat abstracts away all these optimizations from the user who only
sees a vertex program abstraction (SSSP example 
in Figure~\ref{fig:sssp-spmv} and appendix gives an indication of the
effort involved). Hence we are able to get close to the
same performance as native code with much lower programming effort.

\begin{table}[htb!]
  \centering
{\small
  \begin{tabular}{|c|c|}
   \hline
Algorithm & Slowdown compared to \\
& native code in \cite{SIGMOD:2014}\\
\hline
PageRank &  1.15 \\
Breadth First Search &  1.18 \\
Triangle Counting &  2.10 \\
Collaborative Filtering & 0.73 \\
\hline
Overall (Geomean) & 1.20 \\
\hline
  \end{tabular}
}
\caption{Comparison of GraphMat performance to native, optimized code.}
  \label{tab:native-vs-graphmat}
\end{table}

We additionally compare with a recently published SSSP GPU implementation
written in CUDA~\cite{sssp-gpu-ipdps-2014}. The Workfront sweep method described in~\cite{sssp-gpu-ipdps-2014} matches our SSSP implementation. When run on the same Flickr and
Graph500 RMAT Scale 24 graphs, GraphMat is 1.4X and 2.1X slower than the CUDA implementation on an Nvidia GTX 680 GPU, 
in spite of the 3X higher compute and memory bandwidth of the GPU
compared to the system we run on. This shows that
GraphMat utilizes hardware resources more efficiently than the optimized CUDA implementation.

\subsubsection{Scalability}
\begin{figure}[htb]%
\begin{center}
        \subfigure[{PageRank}]
        {
                \label{fig:scale-pagerank}
                \includegraphics[width=0.4\textwidth]{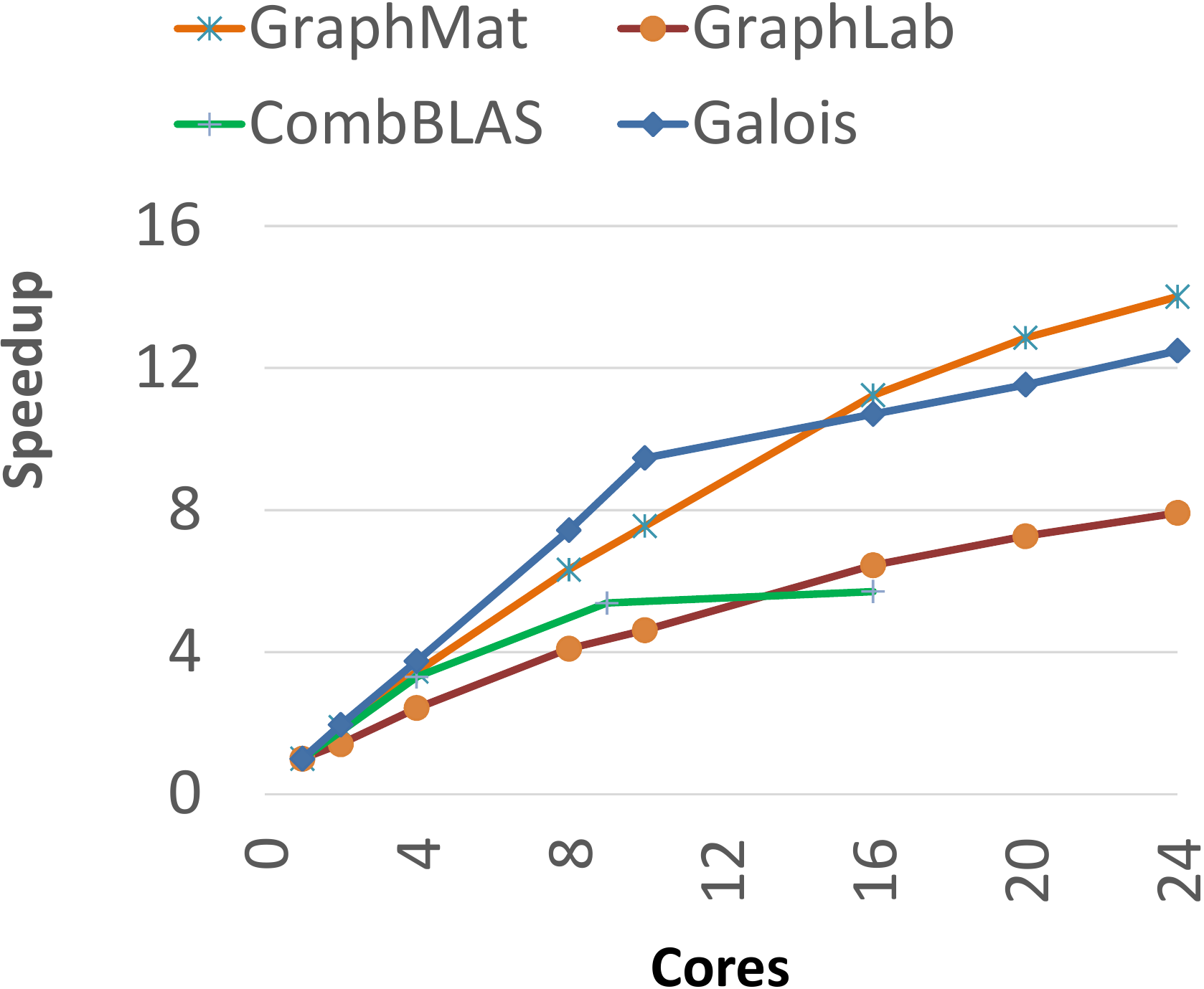}
        }
	\\
        \subfigure[{Single Source Shortest Path}]
        {
                \label{fig:scale-sssp}
                \includegraphics[width=0.4\textwidth]{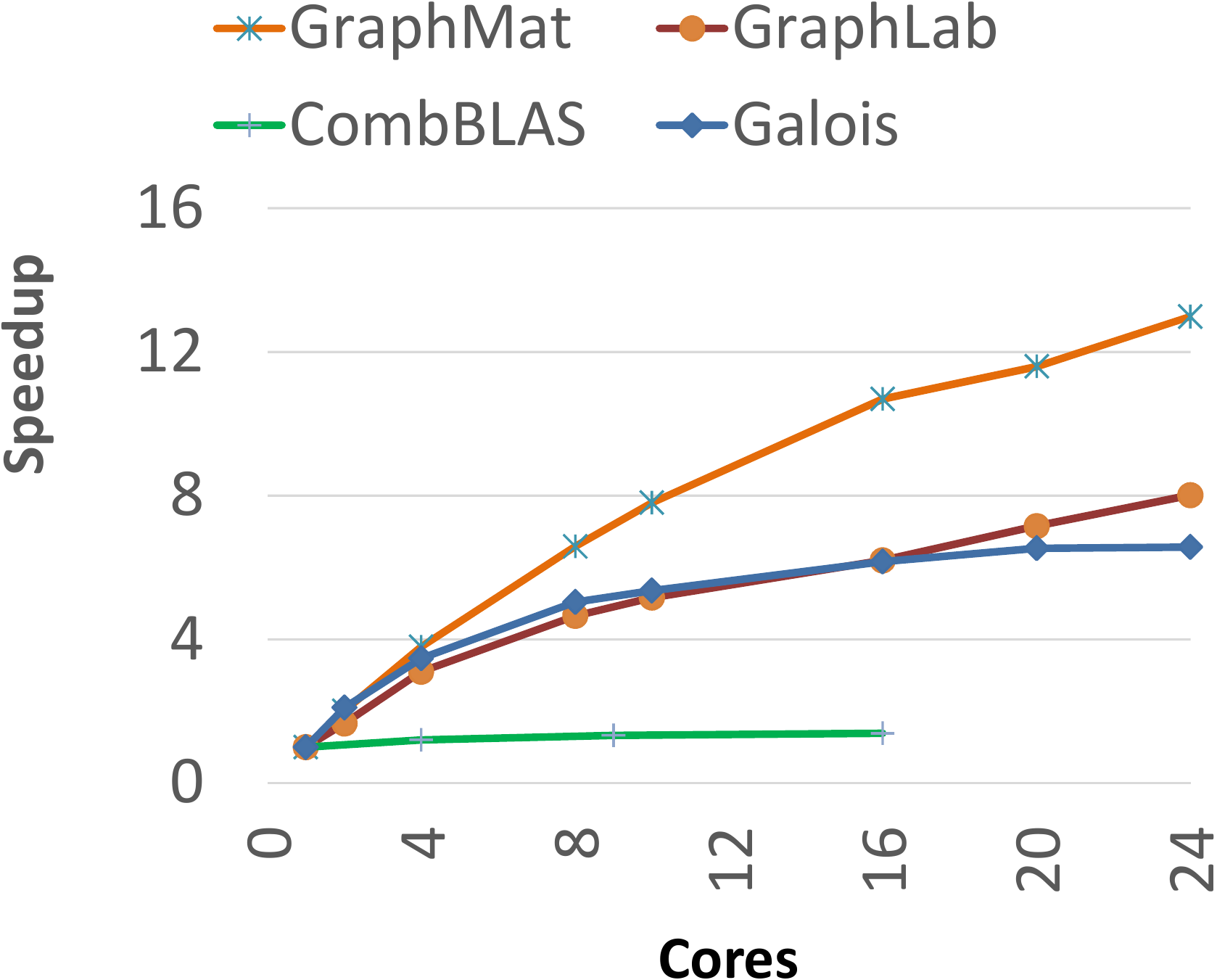}
        }
\caption{\small Scalability of the frameworks using pagerank and single source shortest path algorithms on facebook and flickr datasets respectively.}
\label{fig:scalability-pagerank-sssp}
\end{center}
\vspace*{-0.20in}
\end{figure}

As most performance improvements across recent processor generations
have come from increasing core counts, it has become 
important to consider scalability when choosing
application frameworks as an end-user. 
In this context, we now discuss the scalability of GraphMat and compare it to that of
GraphLab, CombBLAS and Galois. 
Figure~\ref{fig:scalability-pagerank-sssp} shows the scalability
results for two representative applications - Pagerank and SSSP. We
see that no framework scales perfectly linearly with cores, but this is expected
since there are shared resources like memory bandwidth that limit the 
scaling of graph workloads. However, among the frameworks, we can see
that GraphMat scales about 13-15X on 24 cores, while GraphLab,
CombBLAS and Galois only scale about 8X, 2-6X and 6-12X respectively. The trends for other
applications are similar. As a result, on future platforms
with increasing core counts, we expect GraphMat to continue to outperform
GraphLab, CombBLAS and Galois.  

These scaling results do not completely account for the better
performance of GraphMat over GraphLab and CombBLAS; GraphMat
performs better than most frameworks even when all frameworks are run
on a single thread (for example, single-threaded GraphMat is 2-2.5X faster than
CombBLAS and about 8-12X faster than GraphLab). Hence even the baseline for Figure~\ref{fig:scalability-pagerank-sssp} is generally better for GraphMat compared to others. For SSSP alone, Galois has better single thread performance (1.5X) compared to GraphMat as it runs fewer instructions (Section \ref{subsec:perfdiscussion}). However, Galois scales worse than GraphMat for this algorithm. In the next section, we discuss the reasons why
GraphMat outperforms other frameworks.

\subsection{Discussion of performance}
\label{subsec:perfdiscussion}

\begin{figure*}[htb]%
\begin{center}
        \subfigure[{PageRank}]
        {
                \label{fig:pr-emon}
                \includegraphics[width=0.45\textwidth]{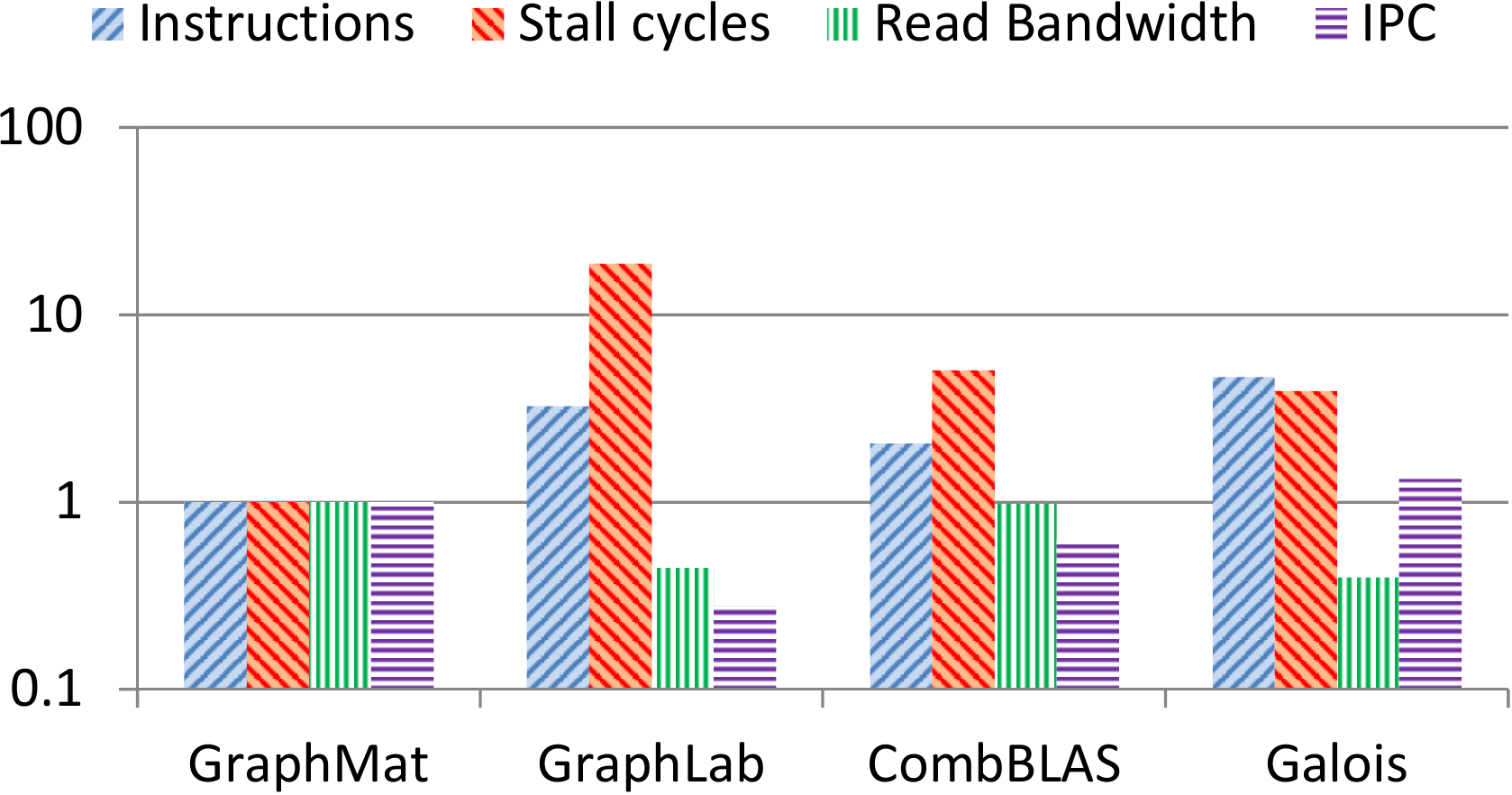}
        }
        \subfigure[{Triangle Counting}]
        {
                \label{fig:tc-emon}
                \includegraphics[width=0.45\textwidth]{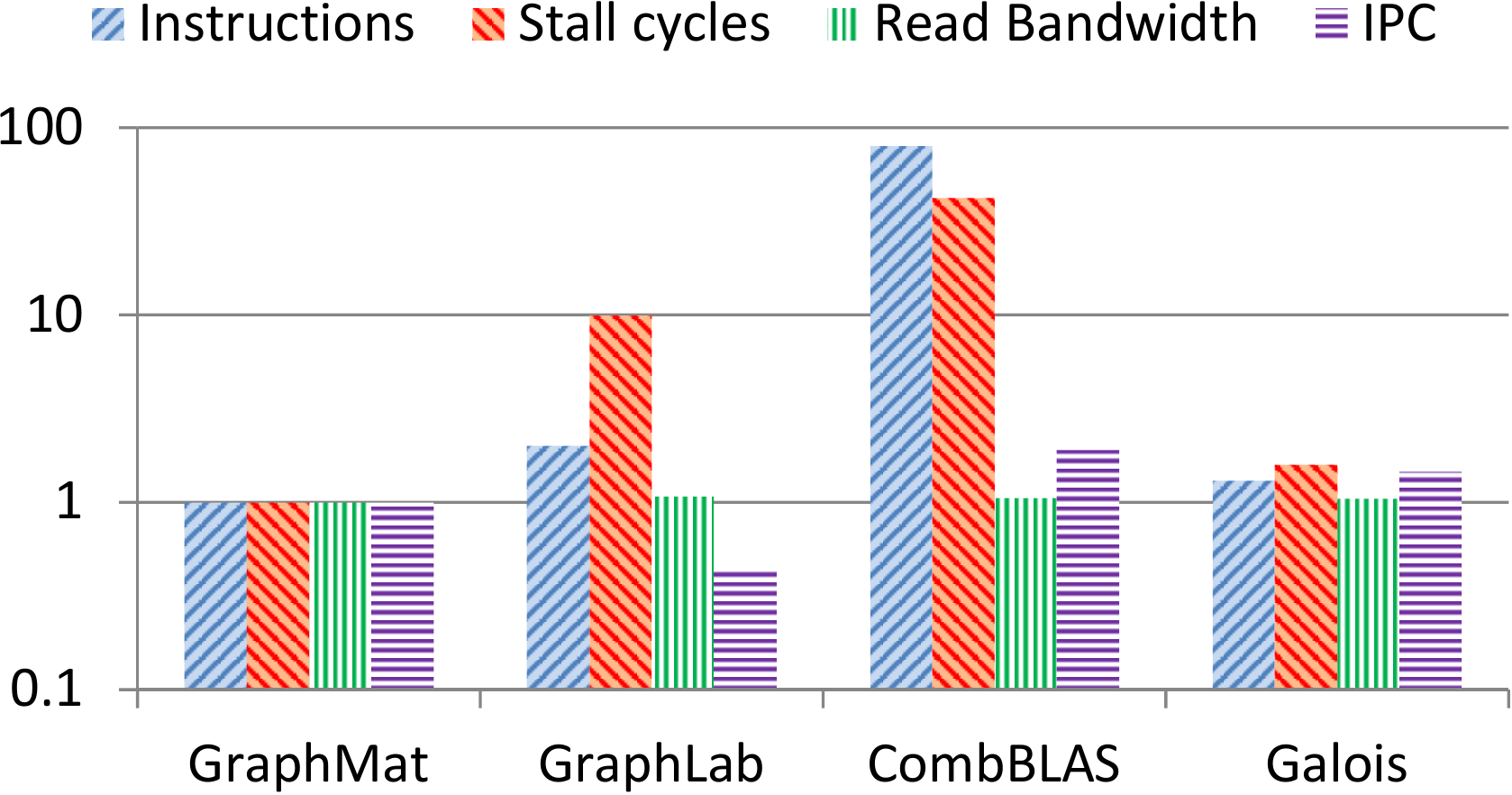}
        }
\\
        \subfigure[{Collaborative Filtering}]
        {
                \label{fig:cf-emon}
                \includegraphics[width=0.45\textwidth]{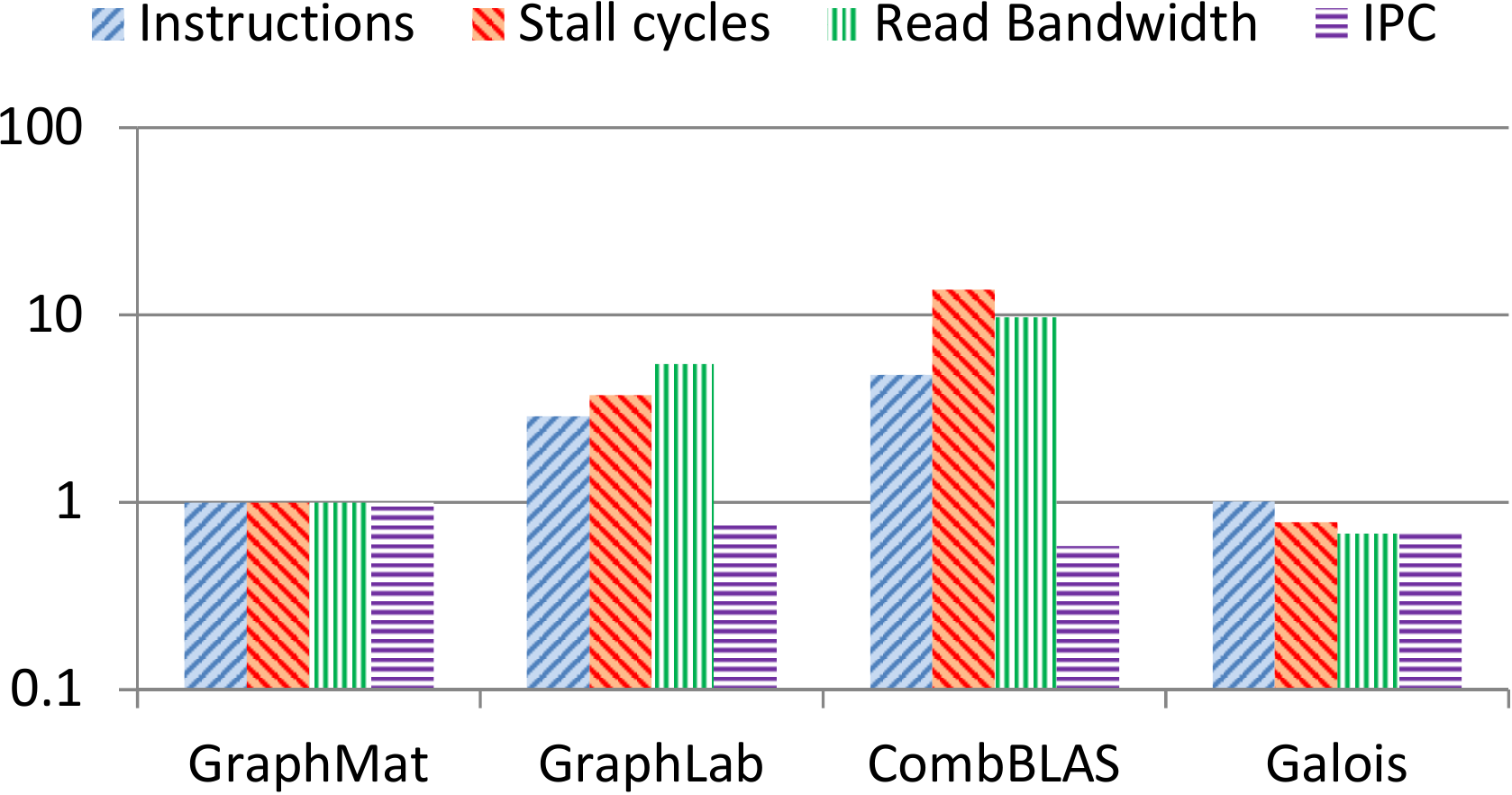}
        }
        \subfigure[{Single Source Shortest Path}]
        {
                \label{fig:sssp-emon}
                \includegraphics[width=0.45\textwidth]{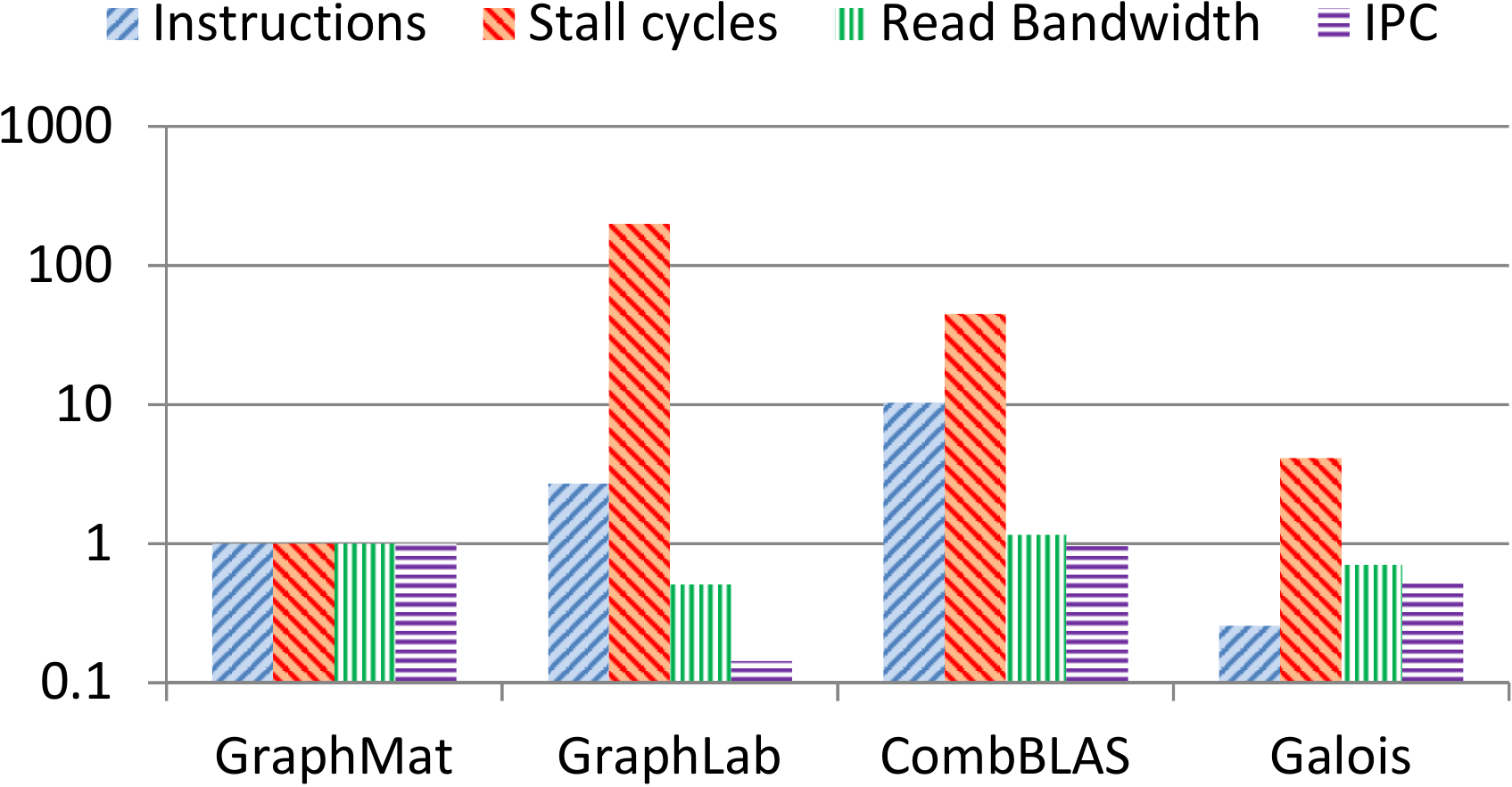}
        }

        \caption{\small Hardware performance counter data for different algorithms
        averaged over all graphs and normalized to GraphMat. The y-axis is in $\log$-scale. Lower numbers are better for instructions and stall cycles. Higher numbers are better for Read bandwidth and IPC.}
        \label{fig:emon}
\end{center}
\vspace*{-0.20in}
\end{figure*}

To understand the performance of the frameworks, we performed a detailed analysis with hardware performance counters. Performance counters are collected for the duration of the application run reported in Figure \ref{fig:single}. This approach of collecting cumulative counters results in better fidelity than sampling based measurements, particularly when the runtimes are small.
Figure \ref{fig:emon}
shows the collected data for four of the applications on all the frameworks. Since graph analytics
operations are mostly memory bandwidth and latency constrained, we focus on counters measuring memory performance. 
For space reasons, we present only the following key metrics that summarize our analysis:
\begin{description}
\item[1. Instructions :] Total number of instructions executed during the test run.

\item[2. Stall cycles :] Total number of cycles CPU core stalled for any reason. Memory related reasons accounted for most of the stalls in our tests.

\item[3. Read Bandwidth :] A measure of test's memory performance. Write bandwidth is not shown since our tests are mostly read-intensive.

\item[4. Instructions per cycle (IPC) :] A measure of test's overall CPU efficiency.
\end{description}
Of these metrics, well-performing code executes fewer instructions, encounters fewer stalls and achieves high read bandwidth and high IPC.

In general, an increase in instruction count and lower IPC indicates overheads in code such as lack of vectorization (SSE, AVX), redundant copying of data and wasted work. Increased stall cycles and reduced memory bandwidth indicates memory inefficiencies which can be remedied through techniques like software prefetching, removing indirect accesses etc. We find that GraphMat is overall at the top (or second best) for most of these indicators.

From Figure \ref{fig:emon}, it is clear that compared to GraphMat, GraphLab and CombBLAS execute significantly more instructions and have more stall cycles. This explains the speedup of GraphMat over both GraphLab and CombBLAS. Even when those frameworks achieve better memory bandwidth than GraphMat (e.g. Collaborative Filtering), the benefits are still offset by the increase in instruction count and stall cycles, implying lots of unnecessary memory loads and wasted work leading to overall slowdown. Galois performs worse than GraphMat for PageRank due to increased instruction and stall cycle count as well. However, Galois performs better than GraphMat on Triangle counting due to better IPC. For Collaborative Filtering, GraphMat has a better IPC and performs better than Galois. For SSSP, Galois uses asynchronous execution (updated vertex state can be read immediately before the end of the iteration) and hence executes fewer instructions, leading to a 1.35X speedup over GraphMat. With GraphMat, the updated vertex state can be read only in the next iteration (bulk synchronous).

We now discuss at a higher level the main reasons why GraphMat performs much better than
the other frameworks. With respect to GraphLab,
GraphMat supports a similar frontend but maps the vertex programs to
generalized sparse matrix operations as described in
Section~\ref{sec:graphmat}. This allows capturing of the global structure
of the matrix and allows for various optimizations including better
load balancing, use of efficient data structures and the use of cache optimizations such as use of global
bitvectors. Moreover, similar operations have been well optimized by the
HPC community and we leverage some of their work~\cite{SPMV}. All
these reasons result in more optimized code than a vertex
program backend like GraphLab can achieve.

On the other hand, CombBLAS uses a similar matrix backend as GraphMat, but
GraphMat still performs about 7X better on average. There are two
primary causes for this. The first is a programming abstraction
reason: 
GraphMat allows for vertex
state to be accessed while processing an incoming message (as
described in Section~\ref{subsec:graphmatmapping}), while CombBLAS
disallows this. There are two algorithms, namely Triangle Counting and
Collaborative Filtering where this ability is very useful both to
reduce code complexity and to reduce runtime as discussed previously in Section \ref{section:genspmv}. 

The second reason for GraphMat to perform better than CombBLAS is a
better backend implementation. For Pagerank, BFS, and SSSP,
the basic operations performed in the backend are similar in GraphMat
and CombBLAS. However, we have heavily optimized our
generalized SPMV backend as described in
Section~\ref{subsec:graphmatopt}. 

Compared to GraphMat, Galois' performance differs by 1.2X. Galois is a sophisticated worklist management system with
support for different problem-specific schedulers \cite{Nguyen:2013:SOSP}, whereas GraphMat is built on top of 
sparse matrix operations. There is not a huge performance difference between the two frameworks on a single node. Extending an efficient task-queue based framework like Galois to other systems (co-processors, GPU,
distributed clusters etc.) however, is a difficult task. In contrast, sparse matrix problems
are routinely solved on very large and diverse systems in the High Performance Computing world. We also note that
GraphMat scales better than Galois with increasing core counts. 
Hence, we believe GraphMat offers an easier and more efficient path to scaling and extending
graph analytics to diverse platforms.

We next describe the performance
impact of the optimizations described in
Section~\ref{subsec:graphmatopt}.

\subsection{Effect of optimizations} 

\begin{figure}[htb]%
\begin{center}
	\includegraphics[width=0.4\textwidth]{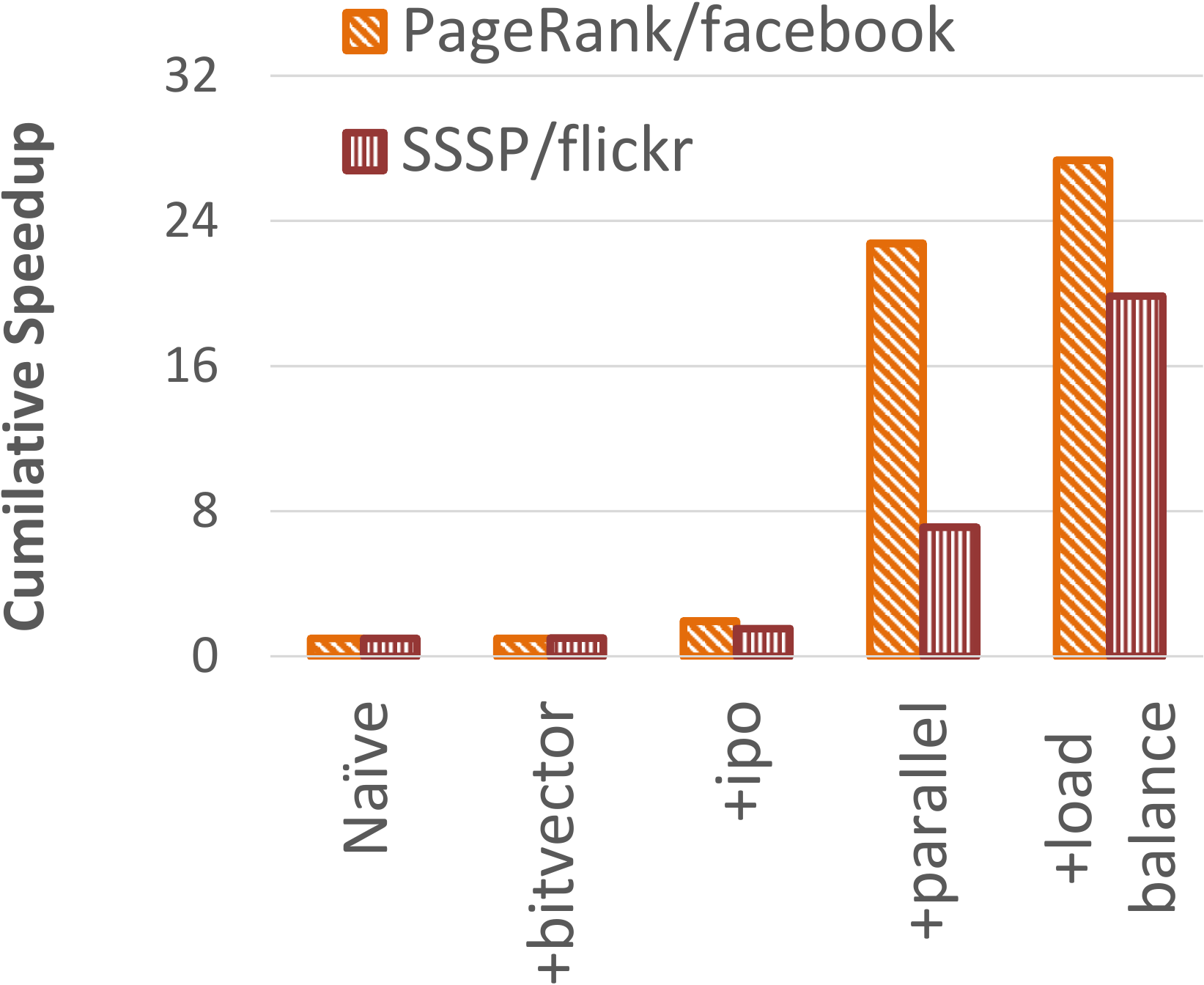}
\caption{\small Effect of optimizations performed on naive implementations of pagerank and single source shortest path algorithms.}
\label{fig:perf-breakdown}
\end{center}
\end{figure}

One of the key advantages of using a
matrix backend is that there are only a few operations that dictate
performance and need to be optimized. In the algorithms described here, most (over 80\%)
of the time is spent in the Generalized SPMV operation as described
in Algorithm~\ref{alg:spmv}. The key optimizations performed to
optimize this operation are described in
Section~\ref{subsec:graphmatopt}. Figure~\ref{fig:perf-breakdown}
shows the performance impact of these four optimizations for Pagerank
(running on the Facebook graph) 
and SSSP (running on Flickr). The first bar shows the baseline naive single threaded code
normalized to 1. Adding bitvectors to store the sparse vectors to
improve cache utilization itself results in a small performance gain.
However, it enables better parallel scalability. Using the compiler
option of -ipo to perform inter-procedural optimization results in a
significant gain of about 1.5X for SSSP and 1.9X for Pagerank. This
third bar represents the best scalar code. 

The fourth and fifth bars
deal with parallel scalability. The
addition of bit-vectors allows for a parallel scalability of about
11.7X and 4.7X on Pagerank and SSSP respectively (without bitvectors,
these numbers were as low as 3.9X and 3.4X on Pagerank and SSSP
respectively). These scalability results are multiplicative with the
gains from ipo and bitvectors, resulting in the fourth bar. Finally,
load balancing optimizations result in a further gain of 1.2X for
Pagerank and 2.8X for SSSP. This results in overall gains of 27.3X and
19.9X from naive scalar code for Pagerank and SSSP respectively. Similar results were
obtained for other algorithms and datasets as well.

From the GraphMat user's perspective, there is no need to understand or optimize the backend in any form. In fact, there is very little performance tuning left to the user (the only tunable ones are number of threads and number of desired matrix partitions). 

To summarize, SPMV operations are heavily optimized and result in better performance for all algorithms in GraphMat. If one considers vertex programming to be productive, there is no loss of productivity in using GraphMat. Compared to matrix programming models, there are huge productivity gains to be had. Our backend optimizations and frontend abstraction choices (such as the ability to read vertex data while processing messages) make GraphMat productive without sacrificing any performance.

\section{Conclusion and Future Work} \label{sec:conclusion}

We have demonstrated GraphMat, a graph analytics framework that utilizes a
vertex programming frontend and an optimized matrix backend in order
to bridge the productivity-performance gap. We show performance
improvements of 1.2-7X when compared to other optimized frameworks such as
GraphLab, CombBLAS and Galois in addition to scaling better on multiple cores.
GraphMat is about 1.2X off the performance of native, hand-optimized code on
average. For users of graph frameworks accustomed to vertex
programming, this provides an easy option for improving performance.
Given that GraphMat is based on SPMV, we expect it to scale well to
multiple nodes. Furthermore, improvements in single node
efficiency translates to fewer nodes used (for a given problem size)
and will lead to better cluster utilization. Our optimizations to the matrix backend can be adopted by other frameworks such as CombBLAS as well, leading to better performance no matter the choice of programming model. Our work also provides a path for array processing systems to support graph analytics through popular vertex programming frontends.

\bibliographystyle{abbrv}
\bibliography{references}

\balance
\begin{appendix} \label{sec:appendix}

We provide the GraphMat source code for Single source shortest path (SSSP) for reference.

\section{SSSP source code}

\begin{lstlisting}[language=C++, keywordstyle=\color{blue}, commentstyle=\color{magenta}, breaklines=true, basicstyle=\scriptsize,
%
%
%
%
%
%
%
%
%
%
	]
//Distance is stored as a single precision floating point number
typedef float distance_type;
distance_type MAX_DIST = FLT_MAX;

class SSSP : public 
GraphProgram<distance_type, distance_type, distance_type> {
//Graph Programs are templatized with 3 types - 
//1. Message type
//2. Processed message type (same as reduced value type)
//3. Vertex property type
//For SSSP, all are distance_type 
   
  public:

  SSSP() {
    order = OUT_EDGES;
	//Performs path traversals only via out-edges
  }
  
  //Send message - Read source vertex property and generate message
  bool send_message(const distance_type& vertexprop, 
                    distance_type& message) const {
    message = vertexprop;
    return true;
  }
  
  //Process message - Read message, edge weight, destination vertex property and update result (processed message)
  void process_message(const distance_type& message, 
                const int edge_weight, 
                const distance_type& vertexprop, 
                distance_type& result) const {
    result = message + edge_weight;
  }

  //Reduction function - reduce two reduced value types a & b into a
  void reduce_function(distance_type& a, 
                       const distance_type& b) const {
    a = std::min(a,b);
  }

  //Apply - Read reduced value and update vertex property
  void apply(const distance_type& reduced_value, 
             distance_type& vertexprop)  {
    vertexprop = std::min(vertexprop, reduced_value);
  }
};

void run_sssp(char* filename, int nthreads) {

  //Declare graph with distance_type as vertex property
  Graph<distance_type> G; 
  //Read file and create 8*nthreads partitions of matrix
  G.ReadMTX(filename, nthreads*8); 
  //set all distances to infinity
  G.setAllVertexproperty(MAX_DIST);
  
  //Create instance of SSSP program
  SSSP _ssspInstance;
  //allocate workspace for GraphMat
  auto workspace = graph_program_init(_ssspInstance, G);
  
  //Source vertex is 6. Set distance of source to 0 and mark it active.
  int v = 6;
  G.vertexproperty[v] = 0;
  G.active[v] = true;

  //Run GraphMat program _ssspInstance on graph G until convergence
  run_graph_program(&_ssspInstance, G, -1, &workspace);
  //G.vertexproperty is an array that holds the correct shortest distances from vertex 6 to all other vertices
  
  //clear workspace
  graph_program_clear(workspace);

}


\end{lstlisting}

\end{appendix}

\end{document}